\documentclass[12pt,a4paper]{article}

\usepackage{setspace}
\usepackage[top=2cm,bottom=2cm,left=2cm,right=2cm]{geometry}
\usepackage[english]{babel}
\usepackage{multirow}
\usepackage{graphicx}
\usepackage[bottom]{footmisc}
\usepackage{amsmath,amsthm,amssymb}
\usepackage{resizegather}
\usepackage{bm}
\usepackage{booktabs}
\usepackage[figuresright]{rotating}
\usepackage{xcolor}
\usepackage{xurl} 
\usepackage{caption}
\usepackage{subcaption}
\usepackage{balance}

\usepackage{hyperref}
\hypersetup{
    colorlinks=true,
    linkcolor=blue,
    filecolor=magenta,      
    urlcolor=cyan,
    pdftitle={From PID to ADRC and back},
    pdfpagemode=FullScreen,
    }

\usepackage{rotating}
\usepackage{pdflscape}

\usepackage{epstopdf}
\AppendGraphicsExtensions{.tif}

\newtheorem{remark}{Remark}

\newcommand{\Vector}[1]{\bm{#1}}  
\newcommand{\Matrix}[1]{\bm{#1}}  
\newcommand{\Transpose}{\mathrm{T}}  
\newcommand{\refEq}[1]{(\ref{#1})}               
\newcommand{\refFig}[1]{Figure~\ref{#1}}           
\newcommand{\refSec}[1]{Section~\ref{#1}}        
\newcommand{\refTable}[1]{Table~\ref{#1}}        

\newcommand{\Revised}[1]{\textcolor{black}{#1}}

\providecommand{\keywords}[1]
{
  \small	
  \textbf{\textit{Keywords---}} #1
}

\title{From PID to ADRC and back: expressing error-based active disturbance rejection control schemes as standard industrial 1DOF and 2DOF controllers}

\author{Momir Stankovic$^1$, He Ting$^{2}$, Rafal Madonski \\
        \small $^{1}$Military Academy, University of Defence, Belgrade, Serbia \\
        \small $^{2}$Energy and Electricity Research Center, Jinan University, Zhuhai, China \\
        \small $^{*}$E-mails: momir\_stankovic@yahoo.com, heting@jnu.edu.cn\\
}

\date{} 

\begin{document}

\maketitle

\begin{abstract}
    In this paper, we uncover a new connection between standard PI/PID controllers and active disturbance rejection control (ADRC), from which we establish formal conditions of equivalence between the two control schemes. Using the equivalence, we devise a step-by-step procedure of transitioning from PI/PID to error-based ADRC. We also show how to go from 1DOF to 2DOF ADRC while retaining a standard 2DOF PI/PID structure. Both procedures facilitate expressing error-based ADRC schemes as standard industrial 1DOF and 2DOF controllers. This allows the designed controller to have the desired characteristic of ADRC (i.e. strong robustness against internal and external uncertainties) while still being expressed in a form that is familiar to industrial practitioners, where PI/PID structures are still the workhorse of modern control systems. The results of the paper ensure backward compatibility of future ADRC-based solutions and foster the adoption of active disturbance rejection-based methods in industrial practice as a viable alternative to standard controllers. To further support the findings, a set of tests is conducted in time and frequency domain, followed by a comparative analysis in FPGA-in-the-loop simulation utilizing a realistic plant model.
\end{abstract}


\keywords{active disturbance rejection control (ADRC), PI/PID controller, 1DOF/2DOF control structure, equivalence conditions, FPGA-in-the-loop simulation}


\begin{flushleft}
    \flushleft{~~~\textbf{\textit{Acronyms\&Abbreviations---}}}
\end{flushleft}
\begin{tabular}{ll}
    1DOF & one degree-of-freedom \\
    2DOF & two degrees-of-freedom \\
    ADRC & active disturbance rejection control \\
    eADRC & error-based ADRC \\
    ESO & extended state observer \\
    EQ & equivalence \\
    FB & feedback \\
    FPGA & field programmable gate arrays \\
    FIL & FPGA-in-the-loop \\
    HDL & hardware description language \\
    PI & proportional integral \\
    PID & proportional integral derivative \\
    PF & pre-filter \\
    TF & transfer function
\end{tabular}


\begin{flushleft}
    \flushleft{~~~\textbf{\textit{Nomenclature---}}}
\end{flushleft}
\renewcommand{\arraystretch}{1.25}{\begin{tabular}{p{0.1\columnwidth} p{0.8\columnwidth}}
    $C_{\textrm{EQ}n}$ & equivalence transfer function for $n\in\left\{1,2\right\}$ \\
    $C_\mathrm{FB}^\mathrm{PI}$ & feedback transfer function of PI controller \\
    $C_\mathrm{FB}^\mathrm{PID}$ & feedback transfer function of PID controller \\
    $C_\mathrm{FB}^\mathrm{A}$ & feedback transfer function of eADRC \\ 
    $C_\mathrm{FB}^{\mathrm{A}n}$ & feedback transfer function for $n\in\left\{1,2\right\}$ \\
    $C_\mathrm{PF}^\mathrm{PI}$ & prefilter transfer function of the PI controller \\
    $C_\mathrm{PF}^\mathrm{PID}$ & prefilter transfer function of the PID controller \\
    $C_\mathrm{PF}^{\mathrm{A}n}$ & prefilter transfer function for $n\in\left\{1,2\right\}$ \\ 
    $G_{\textrm{P}n}$ & plant transfer function for $n\in\left\{1,2\right\}$ \\
    $G_{\textrm{YD}}$ & transfer function of the disturbance-to-output channel \\
    $G_{\textrm{UN}}$ & transfer function of the noise-to-control channel \\
    $G_{\textrm{ER}}$ & transfer function of the reference-to-error channel \\
    $n$ & system relative order
\end{tabular}}

\section{Introduction}

\subsection{Background and motivation}

There are multiple reasons why so many advanced control algorithms were not successfully transitioned from academia to industry. This issue is very complex and part of a larger discussion on bridging the theory vs. practice gap \cite{TheoryGap}, hence it goes beyond the scope of this work. Here, however, we focus on one specific reason, which is the simplicity and intuitiveness of PID's structure. This reason is often attributed as being one of the key factors behind PID dominance in the control area and, at the same time, lack thereof is pointed as the disqualifying drawback of many advanced controllers.

It should be highlighted at this point that it is not our intention to shock the field control engineer, make him/her leave the comfort zone of PID control, and force to absorb an entirely new control concept that one has to learn how to implement and tune. In this work, instead of going against the current and trying to overcome the popularity, availability, and overall prevalence of PID (which might be a futile endeavor), we choose a different path. We take an advanced control solution and devise a full methodology of transitioning that controller from its original form into a PI/PID-like form while retaining its 'advanced' capabilities. We want to address the drawbacks of standard PI/PID controllers (e.g. limited robustness) by injecting new features while retaining (as close as possible) the style of PID's structure and parameters. The justification for this is simple. Through the process of 'hiding advanced controller in PID clothes', we want to improve control performance while keeping the functional simplicity of PI/PID, which will allow engineers to operate such advanced PID-like controller in a simple, straightforward, and familiar manner. In short, we want to put a powerful control algorithm in the hands of floor engineers, and, by mimicking PI/PID, allow them to straightforwardly set up, commission, and tune the controller, all in a short time and without knowledge of the advanced control algorithm.

For the purpose of this work, we select active disturbance rejection control (ADRC) as the example of an advanced controller, for which we will devise, design, and deploy a transition of its structure to a PI/PID form. What is the justification behind selecting ADRC? It is a powerful, highly practical, general-purpose control methodology~\cite{HanTIE,SimADRC}, which can be applied in a variety of ways, depending on control requirements, availability of certain signals, computational capabilities of the target hardware, etc. Since its inception, ADRC has attracted widespread attention among both scholars and practitioners. Starting with its nonlinear version in \cite{HanTIE}, which was later streamlined by its simplified linear form \cite{GaoScaling}, it has found its way into many application domains. The core principle in ADRC is disturbance rejection, namely estimating and canceling total disturbance in real-time. As pointed out in~\cite{HerbstTF}, ADRC is the strongest contender to overcome the theory vs. practice gap. The ``paradigm shift'' in ADRC, mentioned in \cite{Paradigm}, is enabled by a combination of modern control elements with pragmatic, minimal plant modeling, and control loop tuning efforts. Furthermore, it can also be easily equipped with control features that can be useful in engineering practice \cite{Bumplessss}. Finally, the ADRC methodology is now well-grounded and explained in several languages: differential equation \cite{NowickiGeom,NowickiGeom2}, transfer function \cite{HerbstTF,MomirTFadrcFPGA}, and state space \cite{IJCstability}. Recent surveys on ADRC \cite{OverviewESOWenchao} document its development over the years and share successful applications.

The titular expression ``From PID to ADRC'' is a callback to the seminal work \cite{HanTIE}, where the concept of ADRC was first introduced to a larger, English-speaking audience\footnote{Some earlier works on ADRC were mostly in Chinese.}. \Revised{This expression nicely captures the path of early ADRC development, which was mostly focused on trying to position ADRC as an enticing alternative to PID through various analyses, case studies, and even head-to-head comparisons of field data (some overview papers can be found in \cite{MadonskiSuvey,OverviewESOWenchao}). Please notice that the above-mentioned phrase from \cite{HanTIE} is, however, deliberately modified in this article and reads ``From PID to ADRC and back''. As it recently turned out, ADRC has more in common with standard controllers than initially thought \cite{EquivalenceSira,ADRCpiConnect}. It seems that the current development of ADRC has shifted and now goes towards finding connections with classical controllers rather than distancing from them and claiming superiority. Therefore, motivated by the need to support the adoption of ADRC in industrial practice and to ensure backward compatibility of its future applications, we explore those connections in this paper.}


\subsection{Scope, contribution, and structure}

\Revised{The conversion between PID and ADRC has been receiving a lot of attention from the control community lately. The significance of qualitative comparison of PID and ADRC, even though studied to date mostly in the academic sphere, can soon have major practical implications. Even though PID is still the practitioners' usual first choice for designing feedback loops, ADRC continues to emerge as an attractive competitor \cite{PID2ADRCtransit}. It is thus needed to deeply discuss this subject by looking at the currently available body of work and emphasizing its missing points. There are of course several challenges when analyzing the connection between PID and ADRC. One of them is the underlying control philosophy, which is different for PID (passive, error-centric control) and ADRC (active, disturbance-centric control). Another is the tuning methodology, which has to be somehow reconciled as the number of design parameters for the nominal versions of PID and ADRC is different.}

\Revised{Investigating the relation between ADRC and conventional controllers is not a new topic in the scientific literature. There exists a body of work that looked for parameter formulas connecting ADRC with PI/PID \cite{ADRCnormality,PID2ADRCtransit,PIDadrcEquivalence,SiraFreqDomain,adrcPIDinterpretationFirstOrder,adrcPIDinterpretation,SaifUADRC,StateSpacePID}, tried to derive ADRC tuning rules from the PID parameters \cite{FirstOrderPIADRC,PIDADRCtuning,PIDESOddcls,ZNadrc,ADRCpidTuning}, or attempted to interpret classic controllers as disturbance observer-based structures \cite{HubaUltraLocal,HubaPIDDO}.} In this work, however, the scope is focused on a specific version of ADRC, the so-called error-based ADRC (eADRC), which has not been previously considered in terms of its equivalence with PI/PID schemes. Similarly to the most popular output-based ADRC~\cite{GaoScaling}, the eADRC also utilizes a linear observer and a linear state feedback controller, however, its main input signal is not the system output but the feedback error. The reasoning behind the specific selection of eADRC is that it is easily understandable by control practitioners since its structure allows immediate comparability with existing classical control solutions, including PI/PID, which are also error-based (in most instances). More detailed information about the eADRC can be found in~\cite{CTTMomir,Michalekerror,GeneralError,OutputErrorComparison} with some successful applications shown in \cite{IJCreso,eADRCtorsio,eADRCuav,eADRClaser}.

\Revised{Therefore, this paper focuses on formally expressing eADRC as a standard industrial PI/PID controller. Such an equivalence and the procedures associated with it are verified whether they also hold in experimental scenarios, here realized in a hardware setup with a field programmable gate arrays (FPGA) platform. The contributions of this paper are thus as follows.}
    \begin{itemize}
        \item Establishing formal conditions of equivalence between the class of PI/PID controllers and the eADRC scheme.
        \item Derivation of a step-by-step procedure of transitioning from PI/PID to eADRC.
        \item Derivation of a step-by-step procedure of transitioning from one degree-of-freedom (1DOF) to two degrees-of-freedom (2DOF) eADRC while retaining a standard 2DOF PI/PID structure.
        \item \Revised{Reduction of the gap between control system designers and the capabilities of the FPGA hardware by introducing a relatively simple methodology for control algorithm implementation, prototyping, and validation.}
    \end{itemize}
The newly unraveled connection between standard PI/PID controllers and the two formal procedures facilitates expressing eADRC schemes as standard industrial 1DOF and 2DOF controllers. This allows the designed controller to have the desired characteristic of ADRC (i.e. strong robustness against internal and external uncertainties) while being expressed in a form that is familiar to industrial practitioners, where PI/PID structures are still dominant. The results of the paper ensure backward compatibility of future ADRC-based solutions and foster the adoption of active disturbance rejection-based methods in industrial practice as a viable alternative to standard controllers. To the best of our knowledge, such results were not presented in the literature before and contribute to the current body of knowledge. 


\section{Prerequisites: considered control methods}
\label{sec:Prerequisites}

Before we introduce the control techniques considered in this work, let us first define the class of systems that these techniques have to govern. Here we considered the problem of controlling $n$-th order single-input single-output systems that can be expressed as:
\begin{equation}
    y^{(n)}=\sum_{i=0}^{n-1}a_i \cdot y^{(i)}+b \cdot u(t)+d(t),
    \label{eqn:General_nth_order_plant}   
\end{equation}
where $u(t)$ is the control signal, $y(t)$ is the system output, $b$ is the system uncertain input gain, $d(t)$ is the external disturbance, and $a_i$, for $i=0,1,\ldots,n-1$, are the system uncertain internal parameters. Here we primarily focus on first- and second-order systems, i.e. $n\in\left\{1,2\right\}$, as they are most commonly considered in engineering practice.

The considered control objective here is to manipulate the control signal $u(t)$ in a way that the system output $y(t)$ tracks a reference signal $r(t)$ despite the presence of unmodeled external disturbance $d(t)$ and uncertainties related to the system parameters ($a_i$ and $b$).

\begin{remark}
    We have started with the expression in continuous time domain but throughout the article $s$-domain will also be used as it offers controller representation in a compact and easy to implement way. It also aids the frequency domain analysis (to be shown in \refSec{sec:Perform}). We assume, without loss of generality, that the initial values that appear when making Laplace transformation are equal to zero.
\end{remark}

The design of control structures for \eqref{eqn:General_nth_order_plant}, based on PI/PID controllers and eADRC controller, will be recalled in the next subsections \refSec{sec:standCtrl} and \refSec{sec:1dofADRC}, respectively. One will notice that we consider both 1DOF and 2DOF structures. It should be thus reminded that a degree-of-freedom of a control system is defined as the number of closed-loop transfer functions that can be adjusted independently \cite{Horowitzbook}. The design of control systems is a multi-objective problem, hence a 2DOF control system, although more complex, naturally has advantages over a 1DOF control system \cite{2DOFpid}. 

\subsection{PI/PID controllers}
\label{sec:standCtrl}


\subsubsection{2DOF PID}

The 2DOF PID controller for the considered plant model \eqref{eqn:General_nth_order_plant} can be formulated as \cite{Horowitzbook,2DOFpid}:
\begin{align}
    u_\mathrm{PID}(t) &= K_\mathrm{P} \cdot \left[\beta \cdot r(t)-y(t)\right]\notag\\
    &+ K_\mathrm{I} \int_{0}^{t} \left[r(\tau)-y(\tau)\right] d\tau\notag\\ &+K_\mathrm{D}\left(-\frac{dy(t)}{dt}\right),
 \label{eqn:2dof_pid_time_domain}   
\end{align}
with design parameters $K_\mathrm{P}$, $K_\mathrm{I}$, $K_\mathrm{D}$, and $\beta$. 

In industrial applications, it is often beneficial to introduce low-pass filters in the reference and output channels to improve system's step response and reduce its noise sensitivity. Hence, the Laplace transform of \eqref{eqn:2dof_pid_time_domain}, including reference signal filtered by $F_\mathrm{R}(s)$ and system output filtered by $F_\mathrm{Y}(s)$, can take the form\footnote{\Revised{Throughout the paper, to simplify the analysis and to focus on the behavior of the system in the steady-state, we assume that the initial values that appear when making Laplace transformation are equal to zero.}}: 
\begin{align}
    u_\mathrm{PID}(s) &= K_\mathrm{P} \cdot \left[\beta \cdot r(s) \cdot F_\mathrm{R}(s)-y(s) \cdot F_\mathrm{Y}(s)\right]\notag\\
    &+\frac{K_\mathrm{I}}{s} \cdot \left[r(s) \cdot F_\mathrm{R}(s)-y(s) \cdot F_\mathrm{Y}(s)\right]\notag\\
    &+s \cdot K_\mathrm{D} \cdot \left[-y(s) \cdot F_\mathrm{Y}(s)\right].
 \label{eqn:2dof_pid_s_domain}   
\end{align}
The introduced low-pass filters are commonly of first- or second-order, and their coefficients can be considered as additional tuning parameters.



\begin{remark}
    A common engineering practice is only to filter the derivative term in the above equation. In this paper, however, we will use the more general PID form \eqref{eqn:2dof_pid_s_domain}. The transition between the two filtered PID structures can easily be done utilizing the equivalence shown in \cite{matauvsek2011pid}.
\end{remark}

The 2DOF PID transfer function representation from \eqref{eqn:2dof_pid_time_domain} can also be expressed in an alternative, more compact way:
\begin{equation}
    u_\mathrm{PID}(s)=C_\mathrm{FB}^\mathrm{PID}(s) \cdot \left[r(s) \cdot C_\mathrm{PF}^\mathrm{PID}(s)-y(s)\right],
 \label{eqn:2dof_pid_tf_representation}   
\end{equation}
with:
\begin{align}
    C_\mathrm{FB}^\mathrm{PID}(s) &= \frac{K_\mathrm{D} \cdot s^2+K_\mathrm{P} \cdot s+K_\mathrm{I}}{s} \cdot F_\mathrm{Y}(s),\label{eq:PIDFB}\\
    C_\mathrm{PF}^\mathrm{PID}(s) &= \frac{K_\mathrm{P} \cdot \beta  \cdot s+K_\mathrm{I}}{K_\mathrm{D} \cdot s^2+K_\mathrm{P} \cdot s+K_\mathrm{I}} \cdot \frac{F_\mathrm{R}(s)}{F_\mathrm{Y}(s)}.
    \label{eq:PIDPF} 
\end{align}
representing the feedback (FB) and pre-filter (PF) transfer functions, respectively.

The 2DOF PID structure enables separate tuning of disturbance rejection and reference tracking. When the controller is tasked to give a desired attenuation of disturbances, the disturbance rejection performance can be tuned using parameters $K_\mathrm{P}$, $K_\mathrm{I}$, $K_\mathrm{D}$, and filter $F_\mathrm{Y}(s)$. Parameters $\beta $ and filter $F_\mathrm{R}(s)$ can be tuned to give an appropriate reference response.

\subsubsection{1DOF PID}

The 1DOF PID structure, which is the most commonly seen variant of the PID controller, can be obtained from \eqref{eqn:2dof_pid_tf_representation} by simply substituting $C_\textrm{PF}^\textrm{PID}(s)=1$, which yields:
\begin{equation}
    u_\mathrm{PID}(s)=C_\mathrm{FB}^\mathrm{PID}(s) \cdot \overbrace{\left[r(s)-y(s)\right]}^{e(s)},
 \label{eqn:1dof_pid_tf_representation}   
\end{equation}
where $e(s)$ is the tracking error in $s$-domain.


In the case of 1DOF PID controller, contrarily to the commentary under \refEq{eq:PIDPF}, the reference tracking and disturbance rejection performances cannot be tuned separately using parameters $K_\mathrm{P}$, $K_\mathrm{D}$, $K_\mathrm{I}$, and $F_\mathrm{Y}(s)$.

\subsubsection{2DOF and 1DOF PI}

Typically, in the case of first-order systems (i.e. $n=1$), 2DOF PI control structure is utilized, which can be considered as a special case of the 2DOF PID controller but with $K_\mathrm{D}=0$ and $F_\mathrm{Y}(s)=1$, which gives: 
\begin{equation}
    u_\mathrm{PI}(s) = C_\mathrm{FB}^\mathrm{PI}(s) \cdot \left[r(s) \cdot C_\mathrm{PF}^\mathrm{PI}(s)-y(s)\right],
    \label{eqn:2dof_pi_tf_representation}   
\end{equation}
with its pre-filter and feedback transfer functions:      
    \begin{align}
        C_\mathrm{FB}^\mathrm{PI}(s)&=\frac{K_\mathrm{P} \cdot s+K_\mathrm{I}}{s},\label{eq:PIFB}\\ C_\mathrm{PF}^\mathrm{PI}(s)&=\frac{K_\mathrm{P} \cdot \beta \cdot s+K_\mathrm{I}}{K_\mathrm{P} \cdot s+K_\mathrm{I}} \cdot F_\mathrm{R}(s).\label{eq:PIPF} 
    \end{align}

The 1DOF variant of the 2DOF PI controller can be directly obtained from \refEq{eqn:2dof_pi_tf_representation} assuming $G_\textrm{PF}^\textrm{PI}(s)=1$, which gives:
\begin{equation}
    u_\mathrm{PI}(s) = C_\mathrm{FB}^\mathrm{PI}(s) \cdot \overbrace{\left[r(s)-y(s)\right]}^{e(s)}.
    \label{eqn:1dof_pi_tf_representation}   
\end{equation}

\subsection{Error-based ADRC (eADRC)}
\label{sec:1dofADRC}

The error-based ADRC (eADRC) is briefly recalled here with its key design points. For a more detailed derivation, the reader is referred to \cite{GeneralError}.

By incorporating the system dynamics \eqref{eqn:General_nth_order_plant} into the definition of the tracking error $e(t) \triangleq r(t)-y(t)$, one can express the controlled system using its feedback error dynamics:
\begin{equation}
    e^{(n)}=f_\textrm{e}(t)+b_0 \cdot u(t),
    \label{eqn:eADRC_representation_of_general_nth_order_plant}   
\end{equation}
where:
\begin{equation}
    f_\textrm{e}(t) = r^{(n)}-\sum_{i=0}^{n-1}a_i \cdot r^{(i)}+\sum_{i=0}^{n-1}a_i \cdot e^{(i)} + \Delta{b} \cdot u(t)+d(t), \notag
\end{equation}
is the system's resultant generalized (total) disturbance and $\Delta{b}$ represents the uncertainty of the input gain. 

In the eADRC methodology, the control law for the above system is usually defined as:  
\begin{equation}
    u(t) = \frac{1}{b_0} \cdot \left[  \begin{pmatrix}
    \Vector{k}^\Transpose  &  1
    \end{pmatrix} \cdot \Vector{\hat{x}_\textrm{e}} \right],
    \label{eqn:eADRC_CT_SS_Controller_case1}
\end{equation}
where:
\begin{align}
    \Vector{k}^\Transpose &=
    \begin{pmatrix}
       k_1   & \cdots & k_n
    \end{pmatrix}, \notag \\ 
    \Vector{\hat{x}_\textrm{e}} &= \begin{pmatrix}
        \hat{x}_{e1}(t)  &  \cdots & \hat{x}_{en}(t) &  \hat{x}_{e(n+1)}(t)
    \end{pmatrix}^\Transpose \notag \\
    &= \begin{pmatrix}
        \hat{e}(t)  &  \cdots  & \hat{e}^{(n-1)} & \hat{f_{e}}(t)
    \end{pmatrix}^\Transpose, \notag
\end{align}
are the controller gains vector (to be tuned) and state estimation vector, respectively. The latter term can be obtained using an extended state observer (ESO):
\begin{equation}
  \Vector{\dot{\hat{x}}_\textrm{e}}
    = \Matrix{A} \cdot \Vector{\hat{x}_\textrm{e}}(t)
    - \Vector{b} \cdot u(t)
    + \Vector{l} \cdot \left[ e(t) - \Vector{c}^\Transpose \cdot  \Vector{\hat{x}_\textrm{e}}(t) \right],
    \label{eqn:eADRC_CT_SS_Observer_case1}
\end{equation}
\text{for}~$\Matrix{A} = \begin{pmatrix}
        \Matrix{0}^{n \times 1}  &  \Matrix{I}^{n \times n}  \\
        0  &  \Matrix{0}^{1 \times n}
    \end{pmatrix}
    ,\Vector{b} = \begin{pmatrix}
        \Matrix{0}^{(n-1) \times 1}  \\
        b_0  \\
        0
    \end{pmatrix},
    \Vector{c} = \begin{pmatrix}
        1  \\  \Matrix{0}^{1 \times n}
    \end{pmatrix}$,\\
and $\Vector{l} = \begin{pmatrix}
        l_1  &  \cdots  &  l_{n+1}
     \end{pmatrix}^\Transpose$ is the observer gains vector (to be tuned). 
    
To represent the above control observer-based structure in a transfer function form, the Laplace transform of \eqref{eqn:eADRC_CT_SS_Controller_case1} is substituted into \eqref{eqn:eADRC_CT_SS_Observer_case1}, which gives:   
\begin{equation}
    \Vector{\hat{x}}(s)
    = \left(s \cdot \Matrix{I} -  \Matrix{A}_\mathrm{CL} \right)^{-1}
    \cdot \Vector{l}\cdot e(s),
    \label{eqn:eADRC_CT_TF_Transfer_X_case1}
\end{equation}
where $\Matrix{A}_\mathrm{CL}
    =
    \Matrix{A} - \Vector{l} \cdot \Vector{c}^\Transpose + \frac{1}{b_0} \cdot  \Vector{b} \cdot \begin{pmatrix}
     \Vector{k}^\Transpose   & 1 
    \end{pmatrix}$.

By substituting \eqref{eqn:eADRC_CT_TF_Transfer_X_case1}  in the Laplace transform of \eqref{eqn:eADRC_CT_SS_Observer_case1}, and eliminating $\hat{e}(s)$, one arrives at:
\begin{align}
    u(s)=\frac{1}{b_0} \cdot \begin{pmatrix}
        \Vector{k}^\Transpose  &  1
    \end{pmatrix} \left( s \cdot \Matrix{I} - \Matrix{A}_\mathrm{CL} \right)^{-1} \cdot \Vector{l}\cdot e(s),
    \label{eqn:eADRC_CT_TF_Transfer_U}
\end{align} 
which can be rewritten as 1DOF control structure: 
\begin{equation}
    u(s) = C_\mathrm{FB}^\mathrm{A}(s) \cdot e(s). 
    \label{eqn:ADRC_CT_TF_Representation}
 \end{equation}  
The feedback transfer function $C_\mathrm{FB}^\mathrm{A}$ is of a general form :
\begin{equation}
    C_\mathrm{FB}^\mathrm{A}(s)=\frac{1}{b_0} \cdot \begin{pmatrix}
        \Vector{k}^\Transpose  &  1
    \end{pmatrix} \frac{\operatorname{adj}\left( s \cdot \Matrix{I} - \Matrix{A}_\mathrm{CL} \right)}{\operatorname{det}\left( s \cdot \Matrix{I} - \Matrix{A}_\mathrm{CL} \right)} \cdot \Vector{l}.
    \label{eqn:Feedback_tf_eADRC_case1}
 \end{equation}  
For first- ($n=1$) and second-order ($n=2$) plant models it respectively takes the forms:  
\begin{equation}
C_\mathrm{FB}^\mathrm{A1}(s)=\frac{1}{b_0}\cdot \frac{(k_1l_1+l_2)\cdot s+k_1l_2}{s^2+(k_1+l_2) \cdot s},
\label{eqn:Feedback_tf_eADRC_first_order_plant}
\end{equation}
\begin{equation}
C_\mathrm{FB}^\mathrm{A2}(s)=\frac{1}{b_0}\cdot \frac{(k_1l_1+k_2l_2+l_3)\cdot s^2+(k_1l_2+k_2l_3)\cdot s+k_1l_3}{s^3+(k_2+l_1)\cdot s^2+(l_2+k_1+l_1k_2)\cdot s}.
\label{eqn:Feedback_tf_eADRC_second_order_plant}
\end{equation}

From the above derivations, one can notice that this eADRC has 1DOF control structure that only contains the feedback transfer function (similarly to 1DOF PI and 1DOF PID in \refEq{eqn:1dof_pi_tf_representation} and \refEq{eqn:1dof_pid_tf_representation}, respectively). In \cite{CTTMomir}, we have formally shown that the eADRC does not have an imposed pre-filter transfer function in the reference channel (as opposed to the classic output-based ADRC \cite{GaoScaling}). This has certain benefits, for example, it allows the control designer to freely choose if one wants to use a pre-filter and, if yes, what kind of pre-filter.

\section{Main result: establishing conditions of equivalence}
\label{sec:MainResult}

In this section, we uncover a new connection between standard PI/PID controllers and eADRC. By establishing formal conditions of equivalence between the two control schemes, we devise a step-by-step procedure of transitioning from PI/PID to error-based ADRC (\refSec{sec:transPIPIDeADRC}). We also show how to go from 1DOF to 2DOF ADRC while retaining a standard 2DOF PI/PID structure (\refSec{sec:trans1DOF2DOF}). Both procedures facilitate expressing error-based ADRC schemes as standard industrial 1DOF and 2DOF controllers. The results of this section are collectively visualized in \refFig{fig:Visual_transition_PI/PID_eADRC}.

    

\subsection{Transition from PI/PID to eADRC structure}
\label{sec:transPIPIDeADRC}

In this section, we want to find out if the eADRC transfer functions for first-order ($n=1$) and second-order ($n=2$) systems can be represented as PI and PID transfer functions, respectively, multiplied by an equivalence transfer function.


We start with first-order case ($n=1$) and write:
\begin{equation}
    C_\mathrm{FB}^\mathrm{A1}(s)=C_\mathrm{FB}^\mathrm{PI}(s)\cdot C_\mathrm{EQ1}(s),
 \label{eqn:PID_to_eADRC_transition_first_order_plant}
 \end{equation}   
where $C_\mathrm{EQ1}(s)$ is the equivalence transfer function (to be found). After a visual comparison of PI and eADRC structures in \refSec{sec:Prerequisites}, one can deduce that the equivalence transfer function in this case has to be:
\begin{equation}
    C_\mathrm{EQ1}(s)=\frac{1}{\frac{1}{l_2+k_1}\cdot s+1},
    \label{eqn:C_EQ_11}
\end{equation}
with the PI controller parameters set as:
\begin{equation}
 K_\mathrm{P}=\frac{k_1l_1+l_2}{b_0(l_2+k_1)}, \quad K_\mathrm{I}=\frac{k_1l_2}{b_0(l_2+k_1)}.
 \label{eqn:PI_coefficients_for_PID_transition_to_ADRC}
\end{equation} 

In similar manner, we write down the case of second-order plant ($n=2$): 
\begin{equation}
    C_\mathrm{FB}^\mathrm{A2}(s)=C_\mathrm{FB}^\mathrm{PID}(s)\cdot C_\mathrm{EQ2}(s), \label{eqn:PID_to_eADRC_transition_second_order_plant}
 \end{equation}   
and deduce that the equivalence transfer function $C_\mathrm{EQ2}(s)$ in this case has to take the form:
\begin{equation}
 C_\mathrm{EQ2}(s)=\frac{1}{\frac{1}{(k_2l_1+l_2+k_1)}\cdot s^2+\frac{k_2+l_1}{(k_2l_1+l_2+k_1)} \cdot  s+1}\cdot \frac{1}{F_\mathrm{Y}(s)},
 \label{eqn:C_EQ_12}
\end{equation}
with PID controller gains set as:
\begin{align}
    K_\mathrm{P}&=\frac{k_1l_1+k_2l_3}{b_0(k_2l_1+l_2+k_1)},\notag\\ K_\mathrm{I}&=\frac{k_1l_3}{b_0(k_2l_1+l_2+k_1)},\label{eqn:PID_coefficients_for_PID_transition_to_eADRC}\\ K_\mathrm{D}&=\frac{k_1l_1+k_2l_2+l_3}{b_0(k_2l_1+l_2+k_1)}.\notag
\end{align}

The above procedures enable a relatively simple transition from existing PI/PID controller to eADRC control structure by adding the appropriate equivalence transfer function $C_\mathrm{EQ}(s)$ in serial with existing PI/PID controller. 


\subsection{Transition from standard 1DOF eADRC controller to strict 2DOF eADRC controller}
\label{sec:trans1DOF2DOF}

There are two conclusions one can draw when comparing the classic, output-based ADRC (as defined in \cite{GaoScaling}) with the error-based ADRC (as defined in \cite{CTTMomir}):
\begin{itemize}
    \item Output-based ADRC, as seen in \cite{GaoScaling}, is not a \textit{strict} 2DOF\footnote{The term \textit{strict} 2DOF refers to a controller structure consisting of two closed-loop transfer functions that can be adjusted independently.} control structure, which means that although one can identify two transfer functions in the control system scheme, these transfer functions cannot be independently tuned (i.e. tuning of tracking performance and disturbance rejection is intertwined), which is not in line with the definition of the degree-of-freedom given in \cite{Horowitzbook}.
    \item Error-based ADRC is a 1DOF control structure as only one transfer function (related to feedback) can be identified in the control system scheme. There is no transfer function in the reference channel.
\end{itemize}

Therefore, in this section, we want derive a procedure that will show how to transition from standard 1DOF eADRC controller to a 2DOF eADRC controller (with a \textit{strict} 2DOF structure). \Revised{In other words, the idea is to propose 2DOF eADRC structure as a more general controller than 1DOF eADRC,  which allows to obtain the advantages of a 2DOF structure related to the ability to independently tune of the set-point response and  disturbance rejection performance.}



Analyzing the structure of eADRC feedback transfer function for first-order plant \eqref{eqn:Feedback_tf_eADRC_first_order_plant}, one can notice that it can be alternatively represented as a following PI controller feedback transfer function $C_\mathrm{FB}^\textrm{PI}(s)$ with its output additionally filtered with a first-order low-pass filter $F_\mathrm{Y1}(s)$:
\begin{equation}  
    C_\mathrm{FB}^\mathrm{A1}(s)=\frac{K_\mathrm{P}s+K_\mathrm{I}}{s} \cdot F_\mathrm{Y1}(s), 
\label{eqn:transition_eADRC_to_2DOF_eADRC_first_order_plant}
\end{equation} 
where $K_\mathrm{P}$, $K_\mathrm{I}$ are tuned as in \eqref{eqn:PI_coefficients_for_PID_transition_to_ADRC} and the filter has the form $F_\mathrm{Y1}(s)\equiv C_\mathrm{EQ1}(s)$. Next, based on the equivalence between eADRC and 2DOF PI structures (already established in \refSec{sec:transPIPIDeADRC}), the appropriate eADRC pre-filter transfer function can be obtained for the first-order plant model as:  
\begin{equation}
     C_\mathrm{PF}^\mathrm{A1}(s)=\frac{K_\mathrm{P}\beta s+K_\mathrm{I}}{K_\mathrm{P}s+K_\mathrm{I}} \cdot \frac{F_\mathrm{R}(s)}{F_\mathrm{Y1}(s)}=C_\mathrm{PF}^\mathrm{PI}(s) \cdot C_\mathrm{EQ1}^{-1}(s).
 \label{eqn:eADRC_pre-filter_first_order_plant} 
\end{equation}
The two derived transfer functions, related to the pre-filter \refEq{eqn:eADRC_pre-filter_first_order_plant} and the feedback \refEq{eqn:transition_eADRC_to_2DOF_eADRC_first_order_plant}, constitute the desired 2DOF eADRC structure for first-order systems. 

In similar manner, eADRC feedback transfer function  for the second-order plant model can be expressed as:
\begin{equation}  
    C_\mathrm{FB}^\mathrm{A2}(s)=\frac{K_\mathrm{D}s^2+K_\mathrm{P}s+K_\mathrm{I}}{s} \cdot F_\mathrm{Y2}(s), 
\label{eqn:transition_eADRC_to_2DOF_eADRC_second_order}
\end{equation} 
where $K_\mathrm{P}$, $K_\mathrm{I}$, $K_\mathrm{D}$ are tuned as in \eqref{eqn:PID_coefficients_for_PID_transition_to_eADRC} and $F_\mathrm{Y2}(s)$ is a second-order low-pass filter, defined as:
 \begin{equation}  
    F_\mathrm{Y2}(s)=\frac{1}{\frac{1}{k_2l_1+l_2+k_1}\cdot s^2+ \frac{k_2+l_1}{k_2l_1+l_2+k_1}\cdot s+1}.
\label{eqn:transition_eADRC_to_2DOF_eADRC_second_oreder}
\end{equation} 
Therefore, based on the structure of 2DOF PID pre-filter \eqref{eq:PIDPF}, the eADRC pre-filter transfer function can be derived as:  
\begin{align}
     C_\mathrm{PF}^\mathrm{A2}(s) &= \frac{K_\mathrm{P}\beta s+K_\mathrm{I}}{K_\mathrm{D}s^2+K_\mathrm{P}s+K_\mathrm{I}}\cdot \frac{F_\mathrm{R}(s)}{F_\mathrm{Y2}(s)}\notag\\
     &= C_\mathrm{PF}^\mathrm{PID}(s) \cdot C_\mathrm{EQ2}^{-1}(s).
    \label{eqn:eADRC_pre-filter_second_order_plant} 
\end{align}
This, together with \refEq{eqn:transition_eADRC_to_2DOF_eADRC_second_order} constitute the desired 2DOF eADRC structure for second-order systems.

Finally, after the introduced rearrangements, it should be noticed that now adjusting filter $F_\textrm{R}(s)$ and $\beta$ in \eqref{eqn:eADRC_pre-filter_first_order_plant} and \eqref{eqn:eADRC_pre-filter_second_order_plant} enable reference response tuning in the proposed 2DOF eADRC controller without affecting its disturbance rejection performance.

\begin{figure*}[htb!]
   \centering
   \includegraphics[width=0.95\textwidth]{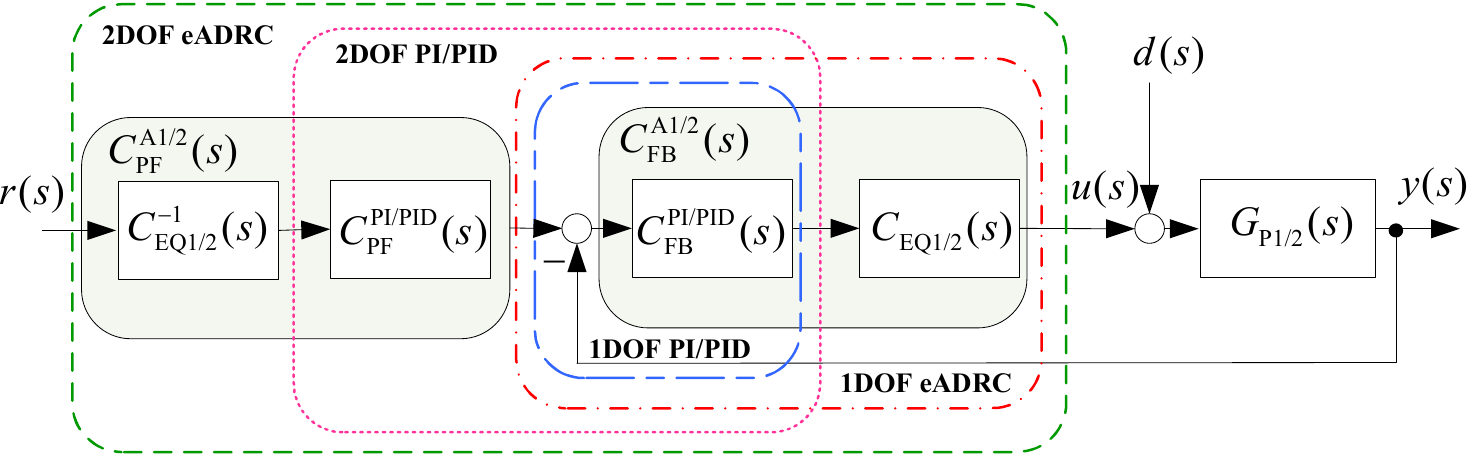}
   \caption{Implementation scheme of PI/PID and eADRC controllers in 1DOF and 2DOF forms. The notation '1/2' corresponds to plant relative order $n=\{1,2\}$. A table containing specific equations for all the considered control schemes can be found in Table~\ref{tab:CheatSheetDetailed} in Appendix~A1.}
   \label{fig:Visual_transition_PI/PID_eADRC}
\end{figure*}

The equivalences between PI/PID and eADRC controllers, obtained in \refSec{sec:transPIPIDeADRC}, enable the transition of 1DOF eADRC to  2DOF eADRC structure based on the 2DOF PI/PID structure. The obtained decoupling of tuning of tracking and disturbance rejection is a highly desirable property in practice, as recently pointed out in \cite{FeedforwardCompensationADRC,2DOFerrorISAT}. The graphical representation of the proposed transitions between PI/PID and eADRC controllers in the 1DOF and 2DOF cases are shown in \refFig{fig:Visual_transition_PI/PID_eADRC}.

\section{Performance analyses in time and frequency domains}
\label{sec:Perform}


In this section, a comparison between considered PI/PID and eADRC structures in both 1DOF and 2DOF cases is performed by analyzing system steady-state and transient performance. \Revised{The comparison is focused on data-driven analysis of the controllers' performance in time and frequency domains, without using equivalent functions. Detailed mathematical analyses of the considered control schemes are omitted here as they have already been covered in the literature, for example in \cite{CTTMomir,GeneralError}.} 

\subsection{Testing methodology}

\Revised{For the purpose of the investigation, two generic plant models are introduced, namely a first-order system model ($n=1$) with time-delay:\footnote{\Revised{Those two specific plant models were also used \cite{CTTMomir}, where the equivalence between eADRC and conventional output-based ADRC was discussed. By using the same examples, one can build on those earlier results, perform comparisons more easily, and expand the topic of the equivalence of ADRC to other control forms.}}}
\begin{equation}
    G_{\textrm{P1}}(s)=\frac{1}{s+1}e^{-0.2\textrm{s}}, 
    \label{eqn:first_order_plant}
\end{equation}
and a normalized second-order system model ($n=2$):
\begin{equation}
    G_{\textrm{P2}}(s)=\frac{1}{s^2+2s+1}. 
    \label{eqn:second_order_plant}
\end{equation}

In order to simplify the tuning of eADRC, the so-called `bandwidth parameterization' is applied. This approach from \cite{GaoScaling} is the most commonly utilized tuning technique for ADRC schemes that has been successfully applied to numerous control scenarios to date \cite{MomirTFadrcFPGA,IJCreso,lakomy2021_DC_DC_Buck}. It enables relatively simple tuning of both controller and observer gains. For the controller, its gains can be obtained by placing closed-loop controller poles at one common location $\lambda = -\omega_\mathrm{CL}$:
\begin{align}
    \left( \lambda + \omega_\mathrm{CL} \right)^n
    = \lambda^{n} + k_n \lambda^{n-1} + \ldots + k_2 \lambda + k_1, 
    \label{eqn:Controller_gains_tuning}
\end{align}  
and in the same manner, observer gains can be obtained placing observer poles at location $\lambda = -k_\mathrm{ESO} \cdot \omega_\mathrm{CL}$:
\begin{align}
    \left( \lambda + k_\mathrm{ESO} \cdot \omega_\mathrm{CL} \right)^{n+1}
    = \lambda^{n+1} + l_1 \lambda^{n} + \ldots + l_{n} \lambda + l_{n+1}, 
    \label{eqn:Observer_gains_tuning}
\end{align}  
where $\omega_\mathrm{CL}$ is the desired  closed-loop system bandwidth and   $k_\mathrm{ESO}$ represents the relative factor between the observer and the control loop bandwidths, both being user-defined design parameters.

To enable a fair comparison between PI/PID and eADRC control schemes, all the considered controller structures are tuned here to have the same robustness index $M_\textrm{s}$ \cite{aastrom2010feedback}:
\begin{align}
    M_\mathrm{s}=\max_{s>0}\left|\frac{1}{1+G_\textrm{FB}(s)G_{\textrm{P}}(s)}\right|.
    \label{eq:Ms}
\end{align}
Therefore, for the derived control structures for first-order plant model \eqref{eqn:first_order_plant}, $M_\textrm{s}=1.55$ is selected by choosing $K_\textrm{P}=1$ and $K_\textrm{I}=2.5$ in case of PI, and in case of eADRC $\omega_\textrm{CL}=2.7$rad/s, $K_\textrm{ESO}=15$, and, assuming complete knowledge of the plant input gain, $b_0=1$. For the controllers for second-order plant model \eqref{eqn:second_order_plant},  $M_\textrm{s}=1.45$ is adopted, and it is achieved by choosing $K_\textrm{P}=30$, $K_\textrm{I}=27$, $K_\textrm{D}=5$, $F_\textrm{Y}(s)=\frac{1}{T_\textrm{f}s+1}$ with $T_\textrm{f}=0.05$s (filter coefficient) in case of PID, and in case of eADRC $\omega_\textrm{CL}=4$rad/s, $k_\textrm{ESO}=7$, and, assuming complete knowledge of the plant input gain, also $b_0=1$.

It should be noted that the corresponding 1DOF and 2DOF structures have identical robustness indices $M_\textrm{s}$, which means that they have the same robustness characteristics, regardless of pre-filter $G_\textrm{PF}(s)$, where $C_\mathrm{PF}(s)$ is either $C_{\textrm{PF}}^{\textrm{PI/PID}}(s)$ or $C_{\textrm{PF}}^{\textrm{A}n}(s)$, depending on the considered plant order ($n\in\left\{1,2\right\}$) and the considered controller type (PI/PID or eADRC). The additional tuning parameters in the 2DOF structures (included in $G_\textrm{PF}(s)$) are set to be the same for both PI/PID and eADRC controllers. The low-pass reference filter is fixed as $F_\textrm{R}(s)=\frac{1}{0.001s+1}$, and in order to show the influence of the parameter $\beta$, its two different values are analyzed (for the first-order plant: $\beta=\left\{0.7;0.3\right\}$ and for the second-order plant: $\beta=\left\{0.75;0.65\right\}$).

\subsection{Steady-state performance analysis}

The steady-state performance analyses of disturbance rejection, noise sensitivity, and reference tracking are conducted in frequency domain, based on the previously defined transfer function representations of the considered PI/PID and eADRC structures shown in \refFig{fig:Visual_transition_PI/PID_eADRC}.  

\begin{figure}
     \centering
     \begin{subfigure}[b]{0.48\textwidth}
         \centering
         \includegraphics[width=\textwidth]{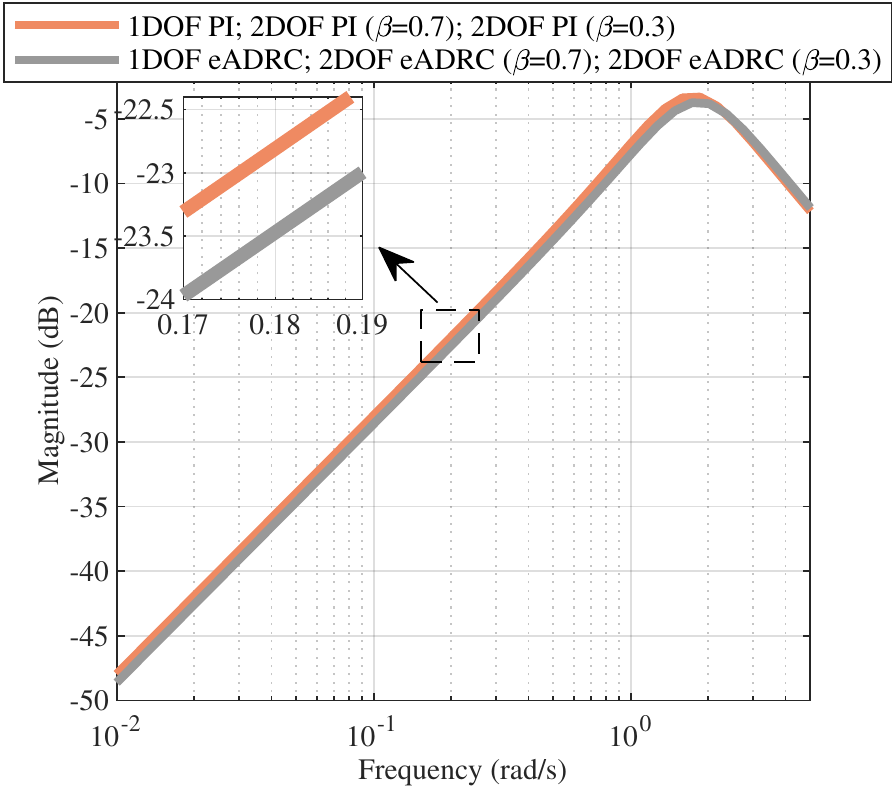}
         \caption{Results for plant model $G_{\textrm{P1}}(s)$. \Revised{Note: corresponding 1DOF and 2DOF structures have the same transfer function $G_\mathrm{YD}(s)$, hence the frequency responses characteristics overlap.}}
         \label{fig:Disturbance_performance_first_orderGP1}
     \end{subfigure}
     \hfill
     \begin{subfigure}[b]{0.495\textwidth}
         \centering
         \includegraphics[width=\textwidth]{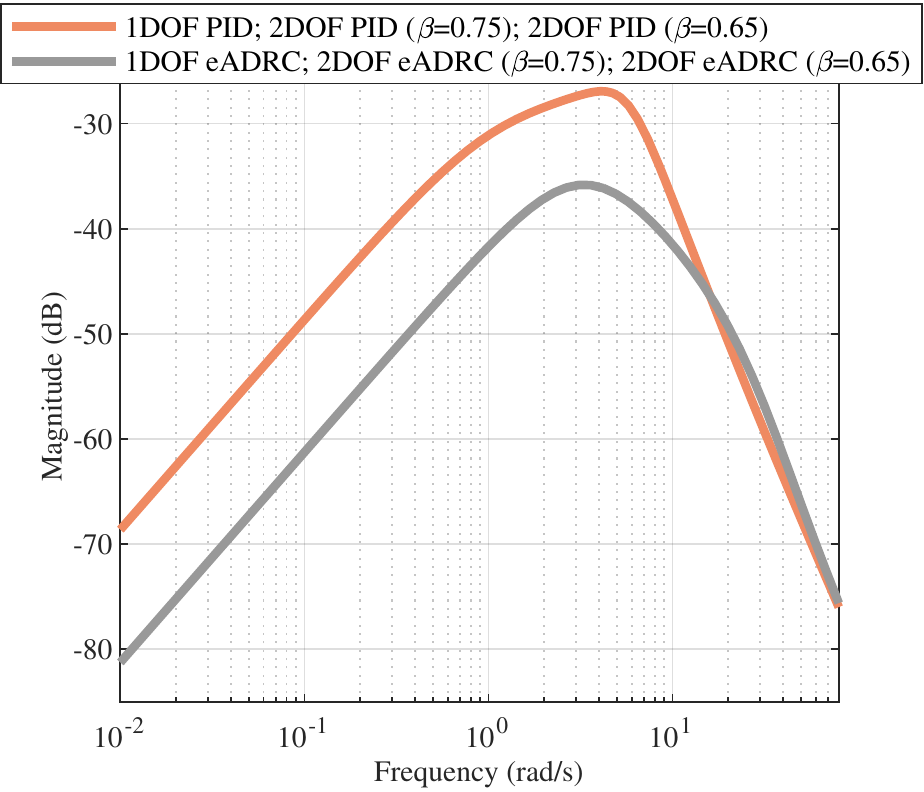}
         \caption{Results for plant model $G_{\textrm{P2}}(s)$. \Revised{Note: corresponding 1DOF and 2DOF structures have the same transfer function $G_\mathrm{YD}(s)$, hence the frequency responses characteristics overlap.}}
         \label{fig:Disturbance_performance_first_orderGP2}
     \end{subfigure}
        \caption{Comparative analyses of steady-state disturbance rejection performances for plant model $G_{\textrm{P1}}(s)$ (a) and $G_{\textrm{P2}}(s)$ (b).}
        \label{fig:Disturbance rejection_steady_state}
\end{figure}

The disturbance rejection performance is analyzed based on frequency response of a following transfer function from disturbance $d(s)$ to output $y(s)$:
\begin{equation}
    G_\mathrm{YD}(s)=\frac{G_{\mathrm{P}n}(s)}{1+C_\mathrm{FB}(s) \cdot G_{\mathrm{P}n}(s)}, 
    \label{eqn:dist_rejection_tf}
\end{equation}
where the specific form of $G_{\textrm{P}n}(s)$ depends on the considered plant order ($n\in\left\{1,2\right\}$), and $C_\mathrm{FB}(s)$ is either $C_{\textrm{FB}}^{\textrm{PI/PID}}(s)$ or $C_{\textrm{FB}}^{\textrm{A}n}(s)$, depending on the considered plant order ($n\in\left\{1,2\right\}$) and the considered controller type (PI/PID or eADRC). The obtained disturbance rejection results are shown in \refFig{fig:Disturbance rejection_steady_state}, jointly for first- and second-order plant.

\begin{figure}
     \centering
     \begin{subfigure}[b]{0.478\textwidth}
         \centering
         \includegraphics[width=\textwidth]{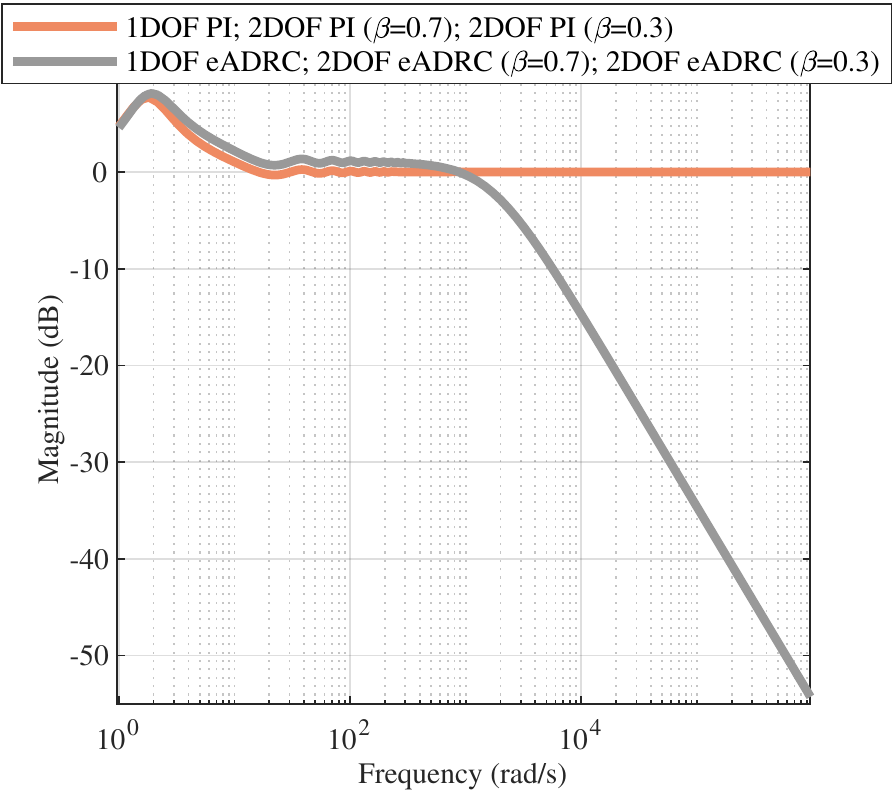}
         \caption{Results for plant model $G_{\textrm{P1}}(s)$ \Revised{Note: corresponding 1DOF and 2DOF structures have the same transfer function $G_\mathrm{UN}(s)$, hence the frequency responses characteristics overlap.}}
         \label{fig:Noise_sensitivity_first_order}
     \end{subfigure}
     \hfill
     \begin{subfigure}[b]{0.495\textwidth}
         \centering
         \includegraphics[width=\textwidth]{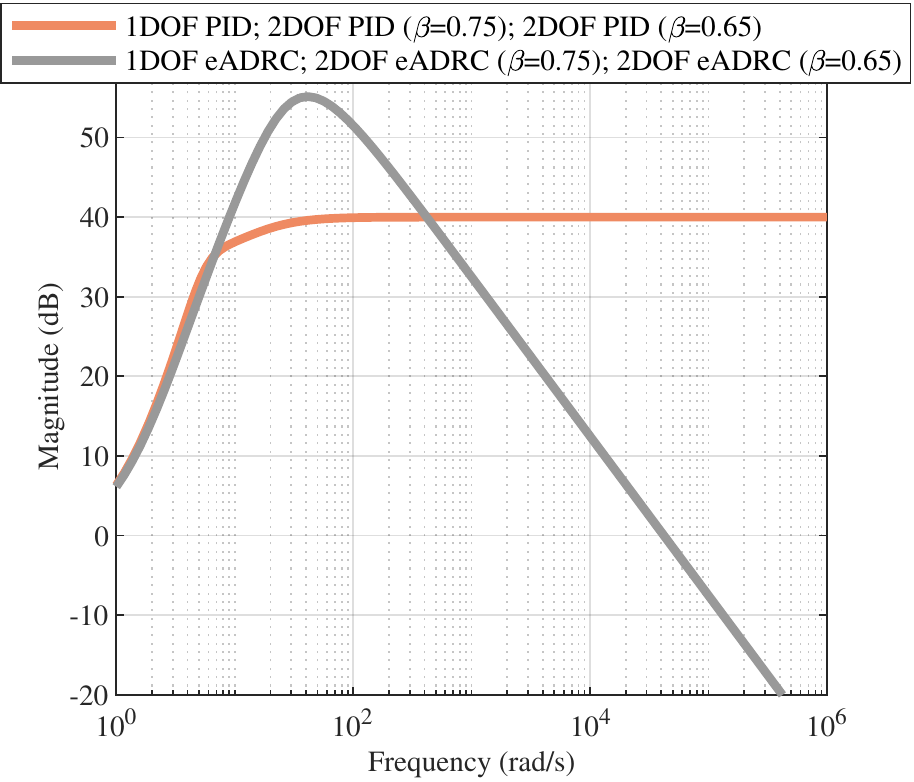}
         \caption{Results for plant model $G_{\textrm{P2}}(s)$ \Revised{Note: corresponding 1DOF and 2DOF structures have the same transfer function $G_\mathrm{UN}(s)$, hence the frequency responses characteristics overlap.}}
         \label{fig:Noise_sensitivity_second_order}
     \end{subfigure}
        \caption{Comparative analyses of steady-state measurement noise sensitivity  for plant model $G_{\textrm{P1}}(s)$ (a) and $G_{\textrm{P2}}(s)$ (b).}
        \label{fig:Noise_sensitivity_steady_state}
\end{figure}

The system noise sensitivity characteristics are obtained based on frequency response of a following transfer function between measurement noise $n(s)$ (in the output signal $y(s)$) and control signal $u(s)$:
\begin{equation}
    G_\mathrm{UN}(s)=\frac{-C_\mathrm{FB}(s)}{1+C_\mathrm{FB}(s) \cdot G_{\mathrm{P}n}(s)}, 
    \label{eqn:noise_sensitivity_tf}
\end{equation}
where the specific form of $G_{\textrm{P}n}(s)$ depends on the considered plant order ($n\in\left\{1,2\right\}$), and $C_\mathrm{FB}(s)$ is either $C_{\textrm{FB}}^{\textrm{PI/PID}}(s)$ or $C_{\textrm{FB}}^{\textrm{A}n}(s)$, depending on the considered plant order ($n\in\left\{1,2\right\}$) and the considered controller type (PI/PID or eADRC). The obtained noise sensitivity results are depicted in \refFig{fig:Noise_sensitivity_steady_state}, collectively for first- and second-order plant. 

One can notice that \eqref{eqn:dist_rejection_tf} and \eqref{eqn:noise_sensitivity_tf} do not contain pre-filter transfer function $C_\textrm{PF}(s)$, which means that appropriate 1DOF and 2DOF controllers have the same disturbance rejection capabilities and noise sensitivity in steady-state, which is additionally confirmed by \refFig{fig:Disturbance rejection_steady_state} and \refFig{fig:Noise_sensitivity_steady_state}. On the other hand, it should be also noted that eADRC controllers enable better disturbance rejection performances than PI/PID controllers (especially in the case of second-order plant) and significantly lower level of noise sensitivity in high frequency range, which is often of interest in the analysis of influence of the measurement noise on system performance.   

In order to compare the performances of steady-state reference tracking, following transfer functions from reference signal $r(s)$ to tracking error $e(s)$ are derived, for 1DOF structure as:
\begin{equation}
    G_\mathrm{ER}(s)=1-\frac{C_\mathrm{FB}(s) \cdot G_{\mathrm{P}n}(s)}{1+C_\mathrm{FB}(s) \cdot G_{\mathrm{P}n}(s)}, 
    \label{eqn:tracking_error_1DOF_tf}
\end{equation}
and for 2DOF structure as:
\begin{equation}
     G_\mathrm{ER}(s)=1-C_\mathrm{PF}(s) \cdot \frac{C_\mathrm{FB}(s) \cdot G_{\mathrm{P}n}(s)}{1+C_\mathrm{FB}(s) \cdot G_{\mathrm{P}n}(s)}, 
    \label{eqn:tracking_error_2DOF_tf}
\end{equation}
where the specific form of $G_{\textrm{P}n}(s)$ depends on the considered plant order ($n\in\left\{1,2\right\}$); $C_\mathrm{FB}(s)$ is either $C_{\textrm{FB}}^{\textrm{PI/PID}}(s)$ or $C_{\textrm{FB}}^{\textrm{A}n}(s)$, depending on the considered plant order ($n\in\left\{1,2\right\}$) and the considered controller type (PI/PID or eADRC); and $C_\mathrm{PF}(s)$ is either $C_{\textrm{PF}}^{\textrm{PI/PID}}(s)$ or $C_{\textrm{PF}}^{\textrm{A}n}(s)$, depending on the considered plant order ($n\in\left\{1,2\right\}$) and the considered controller type (PI/PID or eADRC).

\begin{figure}
     \centering
     \begin{subfigure}[b]{0.49\textwidth}
         \centering
         \includegraphics[width=\textwidth]{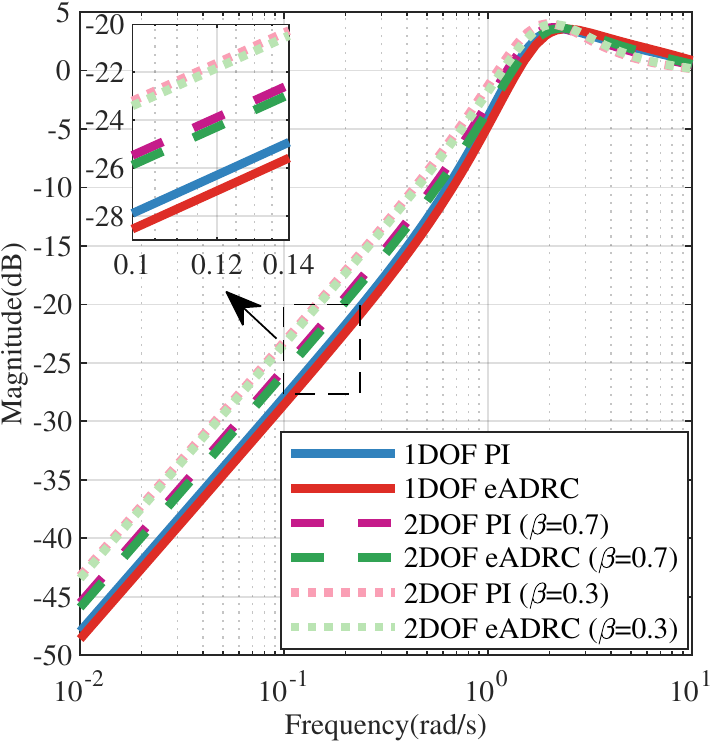}
         \caption{Results for plant model $G_{\textrm{P1}}(s)$}
         \label{fig:Reference_tracking_first_order}
     \end{subfigure}
     \hfill
     \begin{subfigure}[b]{0.49\textwidth}
         \centering
         \includegraphics[width=\textwidth]{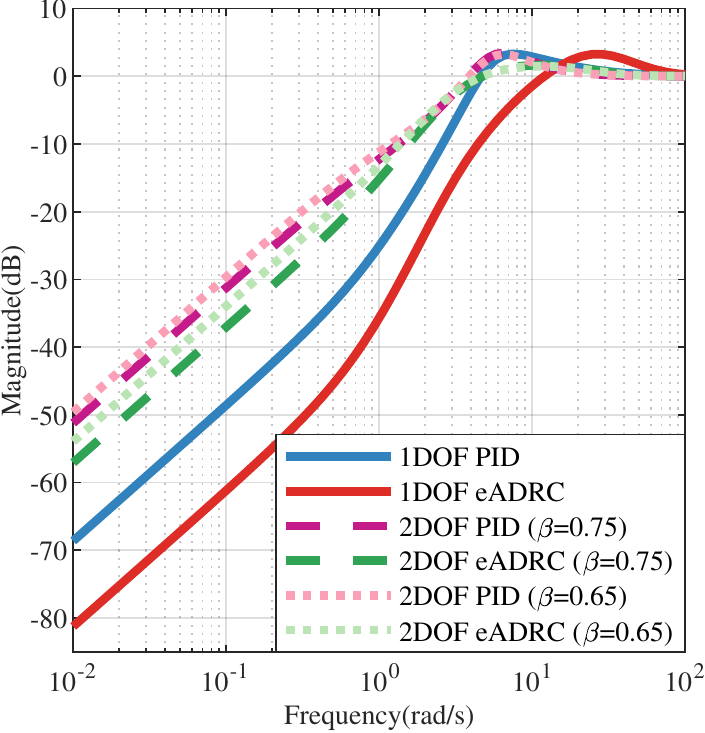}
         \caption{Results for plant model $G_{\textrm{P2}}(s)$}
         \label{fig:Reference_tracking_second_order}
     \end{subfigure}
        \caption{Comparative analyses of steady-state reference tracking performances  for plant model $G_{\textrm{P1}}(s)$ (a) and $G_{\textrm{P2}}(s)$ (b).}
        \label{fig:Reference_tracking_steady_state}
\end{figure}

The frequency responses of $G_\mathrm{ER}(s)$ for the considered controllers are gathered in \refFig{fig:Reference_tracking_steady_state} jointly for first- and second-order plant. It can be observer that 1DOF structures provide better steady-state reference tracking performances than the corresponding 2DOF structures, and that decreasing parameter $\beta$ in 2DOF structure leads to higher reference tracking error in steady-state. Also, comparing eADRC and PI/PID schemes, it is obvious that eADRC achieves better performance in both 1DOF and 2DOF cases.

\subsection{Transient performance analysis}

The transient performances of the considered control structures are analyzed in time-domain using numerical simulations with already established plant models \refEq{eqn:first_order_plant} and \refEq{eqn:second_order_plant}. Simulations are conducted with a step reference signal filtered by a low-pass filter $G_{\textrm{f}}(s)=\frac{1}{(0.01s+1)^2}$. To test the robustness of the considered control algorithms against external disturbance, an unmodeled step disturbance is added from $t \geq 10$s. To test their noise sensitivity, white noise, with power $P_\textrm{n}=1\times 10^{-7}$ and sampling time $T_\textrm{n}=0.001$s, is added to the system output from $t \geq 15$s. 

\begin{figure*}
   \centering
   \includegraphics[width=0.95\textwidth]{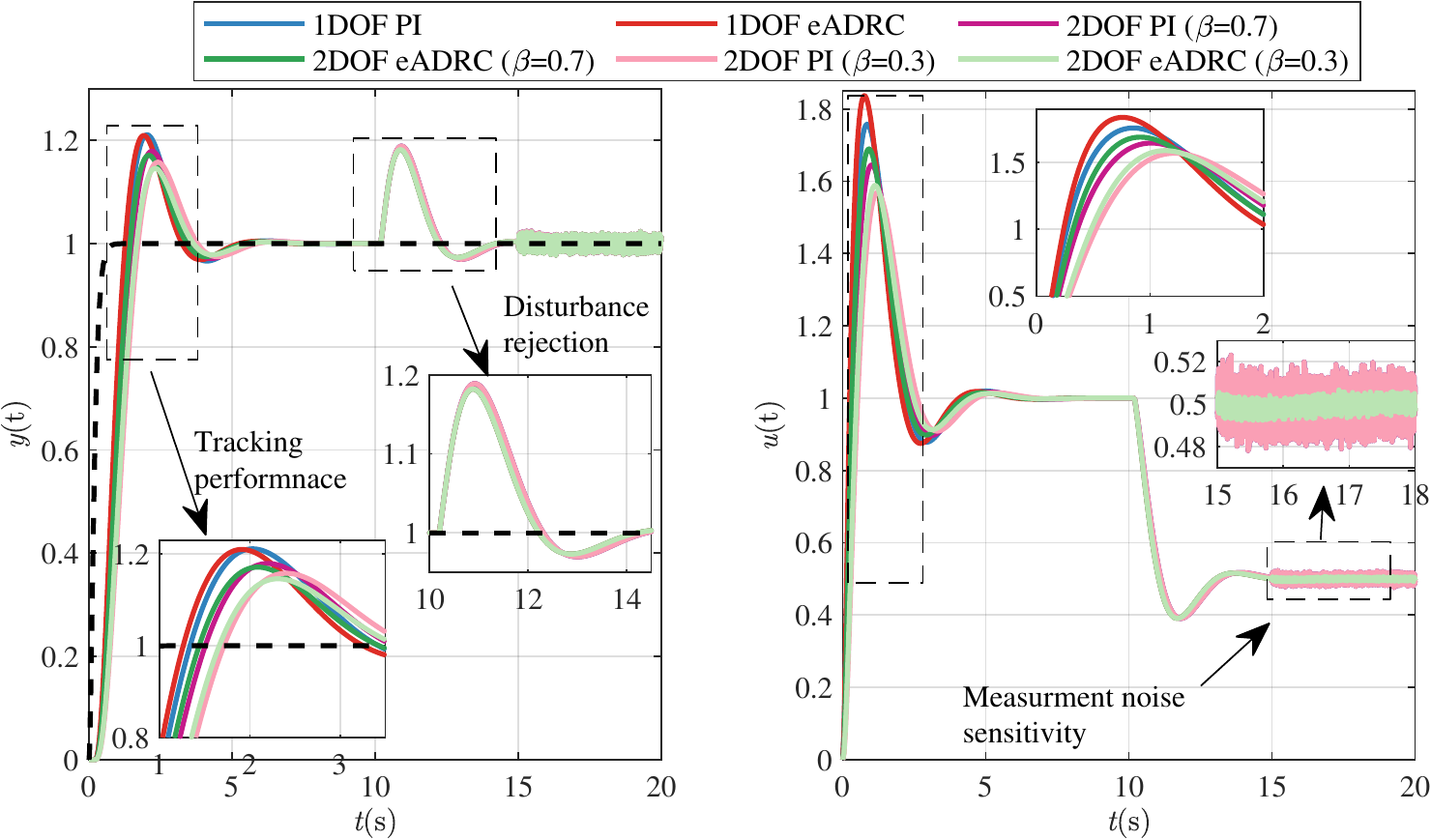}
   \caption{Comparative analyses of transient performances for plant model $G_{\textrm{P1}}(s)$: system outputs (left) and control signals (right). Note: during the disturbance rejection test ($10\textrm{s} \leq t < 20\textrm{s}$) and measurement noise sensitivity test ($15\textrm{s} \leq t \leq 20\textrm{s}$), the signal plots of the 1DOF and 2DOF structures (in both eADRC and PI cases) remain the same, hence are indistinguishable in the plots.}
   \label{fig:Transient_performance_first_order}
\end{figure*}

\begin{figure*}
   \centering
   \includegraphics[width=0.95\textwidth]{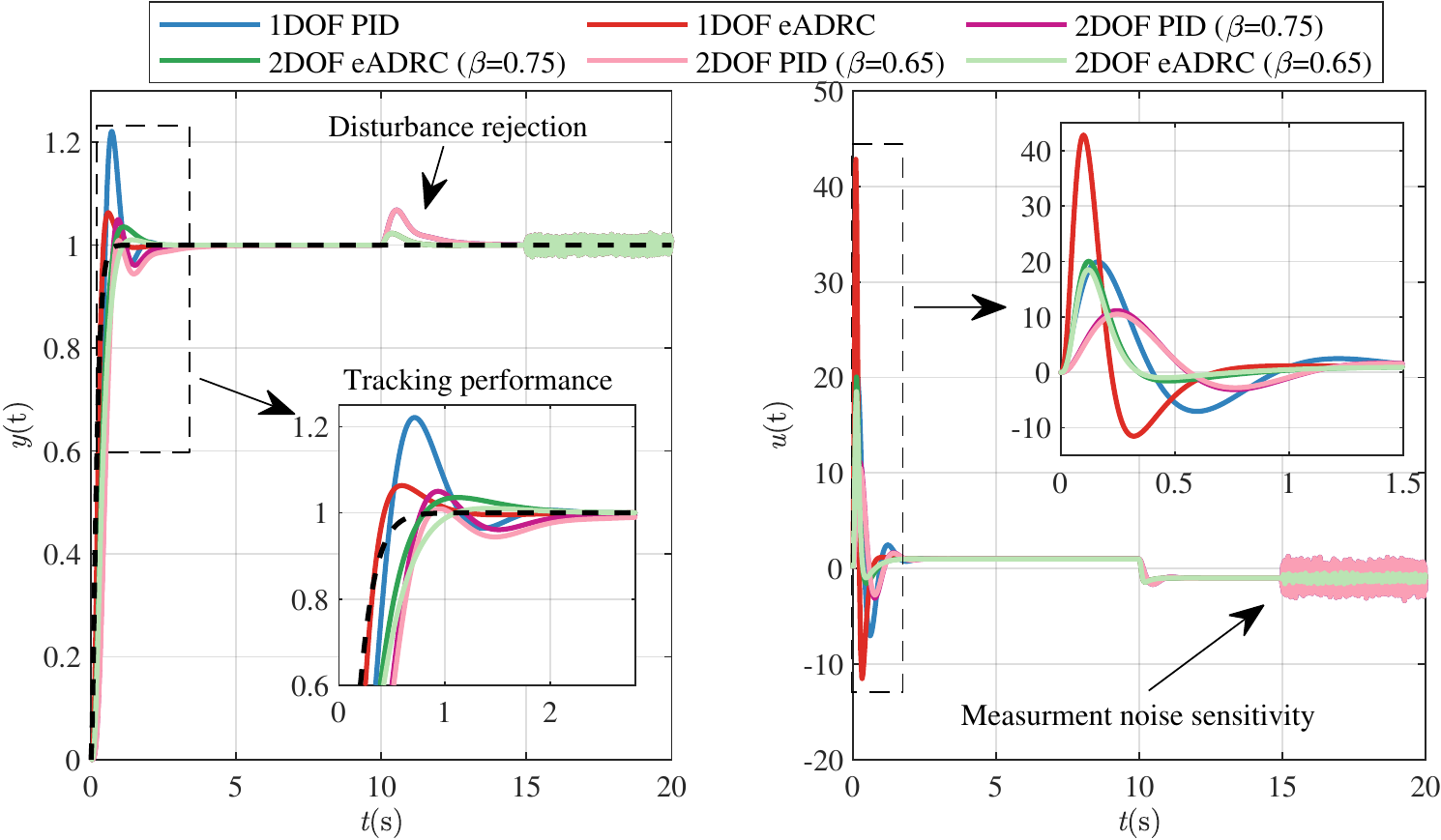}
   \caption{Comparative analyses of transient performances for plant model $G_{\textrm{P2}}(s)$: system outputs (left) and control signals (right). Note: during the disturbance rejection test ($10\textrm{s} \leq t < 20\textrm{s}$) and measurement noise sensitivity test ($15\textrm{s} \leq t \leq 20\textrm{s}$), the signal plots of the 1DOF and 2DOF structures (in both eADRC and PI cases) remain the same, hence are indistinguishable in the plots.}
   \label{fig:Transient_performance_second_order}
\end{figure*}

The obtained systems responses and generated corresponding control signals are presented in~\refFig{fig:Transient_performance_first_order} for the first-order plant, and in~\refFig{fig:Transient_performance_second_order} for the second-order plant. From these figures, it can be noticed that in all the simulated scenarios, eADRC controllers achieved better reference tracking than PI/PID controllers, and that 2DOF structures provide responses with lower overshoots and values of the control signal than 1DOF structures. It can be also seen that the system response and the associated control signal peak can be shaped by adjusting parameter $\beta$, where its lower value means lower overshoot and lower control signal peak. Also, as expected, it is evident that shaping the reference response does not affect the disturbance rejection performance and noise sensitivity, which confirms the previously made conclusion resulting from~\refEq{eqn:dist_rejection_tf} and~\refEq{eqn:noise_sensitivity_tf}.

Therefore, it can be concluded that the proposed 2DOF eADRC controller, derived based on 2DOF PI/PID controller, in comparison with the standard 1DOF eADRC controller, provides the additional ability to adjust the transient reference response, while retaining the same noise sensitivity level as well as disturbance rejection performance in both steady-state and transient stages.     
    
\section{FPGA-in-the-loop simulations}
\label{sec:FPGA}

\subsection{Lab setup and testing methodology}

\begin{figure*}[htb!]
   \centering
   \includegraphics[width=0.75\textwidth]{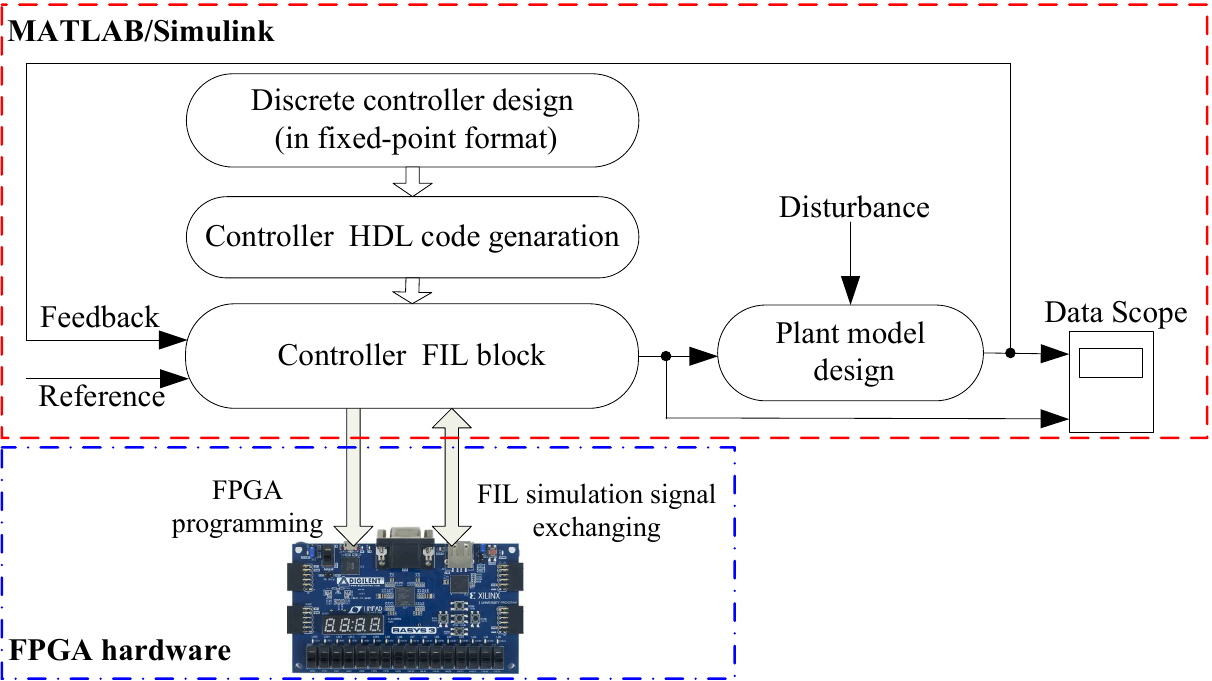}
   \caption{Methodology of the performed FPGA-in-the-loop simulation tests.}
   \label{fig:FIL_concept}
\end{figure*}

\Revised{The goal of this section is to verify whether the discovered equivalence and procedures hold under experimental test. Due to, FPGA-in-the-loop (FIL) validation tests of the considered control algorithms are performed. FPGA-based implementations can be consider as an alternative to commonly used microprocessor-based
implementations for certain applications, especially when fast signal processing or control of more parallel
processes is required \cite{MomirTFadrcFPGA, stankovic2018optimized}. The reasons for the less popular use of FPGA hardware are often the lack of experience
of control designers and engineers in VHDL programming and FPGA implementation. To reduce the gap between control system designers and the capabilities of the FPGA hardware, a relatively simple FIL test methodology, presented in \refFig{fig:FIL_concept}, for control algorithm implementation,
prototyping, and validation, based on low cost FPGA chip and MATLAB/Simulink software, is proposed.} The controller hardware description language (HDL) code is generated using MATLAB/Simulink HDL Code Generation toolbox, based on the controller's discrete form realized in a fixed-point representation format of the controller's signals and coefficients. Then, following MATLAB/Simulink FIL wizard, the controller FIL block is formed by choosing the desired FPGA target platform; in our case, a low-cost Basys~3 board with Xilinx Artix~7 FPGA chip XC7A35T is used. The generated controller FIL block is included in the control loop of the plant model realized in MATLAB/Simulink. By starting the simulation, the controller is implemented in the FPGA chip, and the simulation is executed, such that the controller FIL block calculations are performed on the real FPGA hardware. The FIL simulation signal exchanging process is assured by a micro-USB cable that connects computer USB port and a JTAG port on the FPGA board.



\begin{figure}[htb!]
     \centering
     \begin{subfigure}[b]{0.49\textwidth}
         \centering
         \includegraphics[width=\columnwidth]{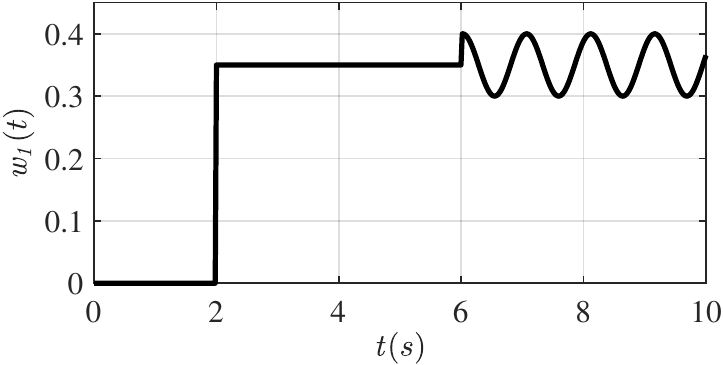}
         \caption{External disturbance in 'Scenario I'}
         \label{fig:disturbance_FIL_1}
     \end{subfigure}
     \hfill
     \begin{subfigure}[b]{0.49\textwidth}
         \centering
         \includegraphics[width=\columnwidth]{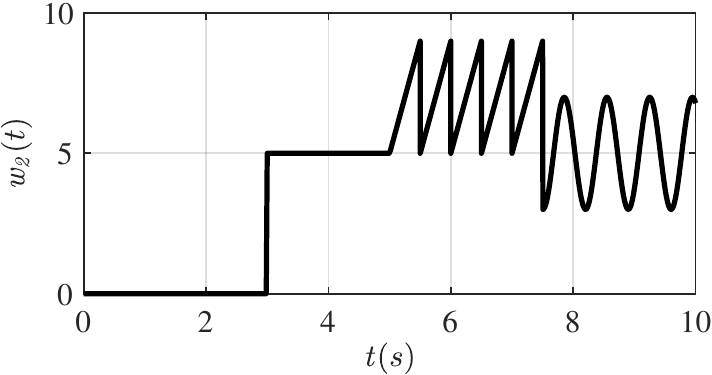}
         \caption{External disturbance in 'Scenario II'}
         \label{fig:disturbance_FIL_2}
     \end{subfigure}
        \caption{\Revised{User-defined applied external disturbance. To make the test case realistic, the information about the acting disturbance is not utilized anywhere in the control design, as to mimic a real-world scenario where the external disturbance would be unknown.}}
\end{figure}

To comparatively analyze performances of the considered control methods, two following scenarios are prepared:
\begin{itemize}
    \item 'Scenario I' --- In this test, the comparative performance analysis of output voltage reference tracking is performed using a highly uncertain model of a DC-DC buck power converter, controlled by 1DOF PID controller \eqref{eqn:1dof_pid_tf_representation} (with first-order filter $F_\textrm{Y}(s)=\frac{1}{T_\textrm{f} \cdot s+1}$), 1DOF eADRC controller \eqref{eqn:Feedback_tf_eADRC_second_order_plant}, and 1DOF eADRC controller (obtained from 1DOF PID by adding equivalence transfer function \eqref{eqn:C_EQ_12}). To further complicate the control task and to check the robustness of the applied control, the governed system is additionally subjected to an unmodeled, nonlinear load disturbance (depicted in \refFig{fig:disturbance_FIL_1}). Details of this scenario are described in \refSec{sec:FILscenario1}.
    \item 'Scenario II' --- In this test, the comparative performance analysis of motor shaft angular speed reference tracking is performed using a highly uncertain model of DC motor, controlled by 1DOF and 2DOF eADRC controllers. To further complicate the control task and to check the robustness of the applied control, the governed system is additionally subjected to an unmodeled, nonlinear load disturbance (depicted in \refFig{fig:disturbance_FIL_2}). Details of this scenario are described in \refSec{sec:FILscenario2}.
\end{itemize}

\subsection{FIL simulation of 'Scenario I'}
\label{sec:FILscenario1}

The used DC-DC buck power converter model is expressed as a following second-order system \cite{lakomy2021_DC_DC_Buck}: 
\begin{equation}
    \ddot{y}_1=-\frac{1}{C \cdot R} \cdot \dot{y}_1-\frac{1}{C \cdot L} \cdot y_1(t)+\frac{V_\mathrm{in}}{C \cdot L} \cdot \left[u_1(t)+w_1(t)\right], 
    \label{eqn:DC-DC_buck_convereter}
\end{equation}
where $y_1(t)$ represents the system output voltage, $u_1(t)$ is the control signal, $R=50\Omega $ is the load resistance, $C=0.001$F is the filter capacitance, $L=0.01$H is the filter inductance, $V_\mathrm{in}=20$V is the input DC voltage source, and $w_1(t)$ represents the user-defined external perturbation (shown in \refFig{fig:disturbance_FIL_1}).

Applying the Euler approximation and substituting $s=(z-1)/T_\textrm{s}$ in \eqref{eq:PIDFB}, \eqref{eqn:Feedback_tf_eADRC_second_order_plant} and \eqref{eqn:C_EQ_12}, the discrete forms of the considered 1DOF PID controller, 1DOF eADRC controller, and the equivalence transfer functions become: 
\begin{align}
    C_\mathrm{FB}^\mathrm{PID}(z)&=\frac{K_\mathrm{D}z^2+(K_\mathrm{P}T_\textrm{s}-2K_\mathrm{D})z+K_\mathrm{D}-K_\mathrm{P}T_\textrm{s}+K_\mathrm{P}T^2_s}{T_\textrm{f}z^2+(T_\textrm{s}-2T_\textrm{f})z+T_\textrm{f}-T_\textrm{s}}
 \label{eqn:PID_z_domain}, \\
    C_\mathrm{FB}^\mathrm{A2}(z)&=\frac{T_\textrm{s}}{b_0} \cdot \frac{M_2z^2+M_1z+M_0}{(z^2+N_1z+N_0)(z-1)},
 \label{eqn:eADRC_z_domain} \\
    C_\mathrm{EQ2}(z)&=\frac{L_1z+L_0}{P_2z^2+P_1z+P_0},
 \label{eqn:C_EQ2_z_domain} 
\end{align}
respectively, where $T_\textrm{s}$ is the sampling time and the coefficients of \eqref{eqn:eADRC_z_domain} and \eqref{eqn:C_EQ2_z_domain} are given in Appendix~A2. 

\begin{remark}
    Although not the case here for the considered FPGA board, but one may want to consider checking other, more efficient discrete implementations of eADRC for applications with low sampling frequencies \cite{CTTHerbst,HerbstMinumum,HerbstCTTdiscrete}.
\end{remark}

\begin{figure*}[htb!]
   \centering
   \includegraphics[width=.9\textwidth]{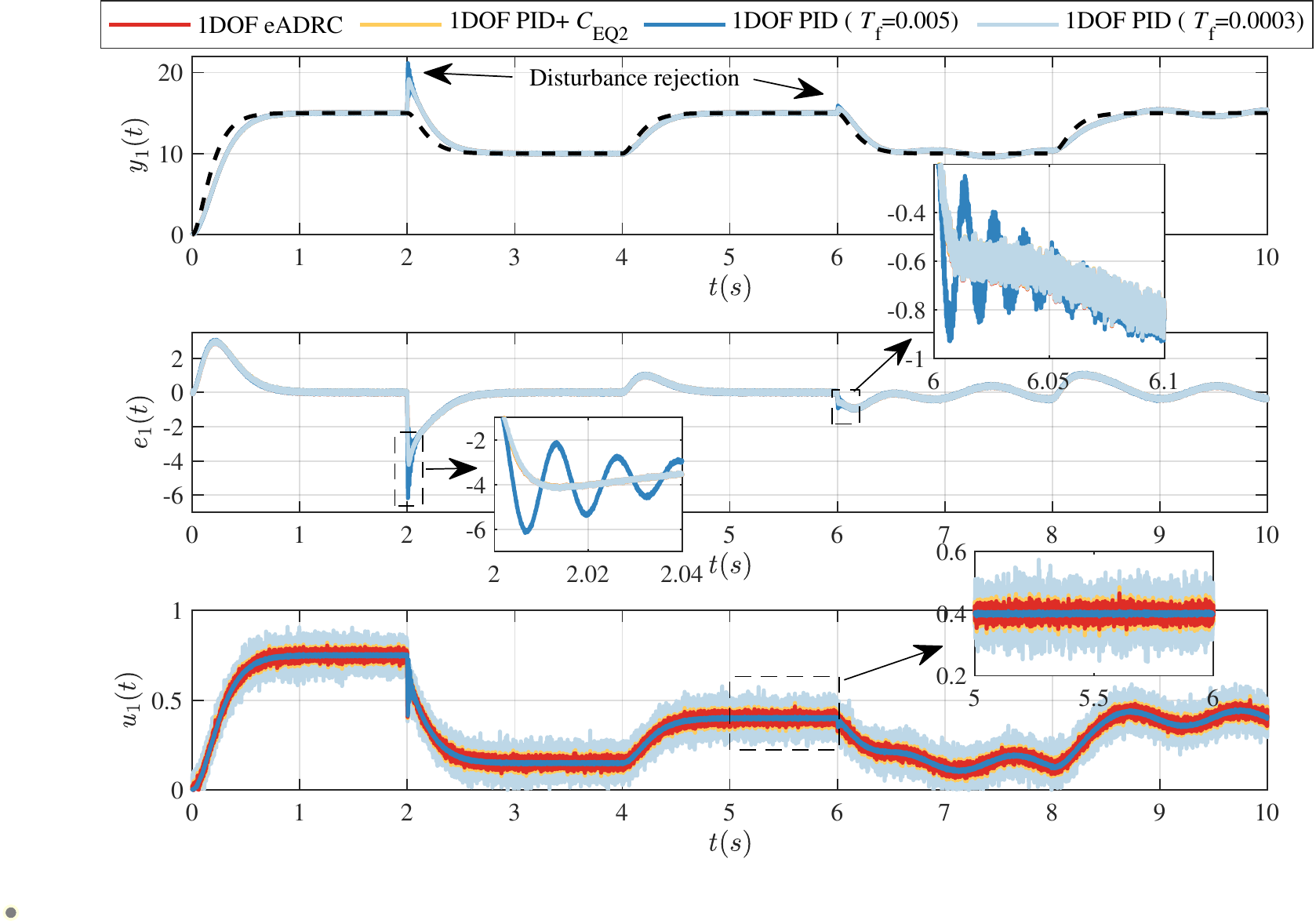}
   \caption{'Scenario I': results of FPGA-in-the-loop simulation with systems outputs (top); tracking errors (middle); control signals (bottom).}
   \label{fig:Results_FIL_1}
\end{figure*}

The eADRC controller is tuned by choosing $\omega_\textrm{CL}=45$rad/s, $k_\mathrm{ESO}=45$, and, assuming complete knowledge of the plant gain, $b_0=V_\mathrm{in}/CL$. The PID controller gains are set based on eADRC parameters using \eqref{eqn:PID_coefficients_for_PID_transition_to_eADRC}, and two different values of the filter parameter $T_\textrm{f}=\left\{0.005,0.0003\right\}$, are analyzed. The considered discrete controllers are realized in high precision 64-bit fixed-point format, with the sampling time $T_\textrm{s}=0.0001$s. The control objective is to track a reference signal being a square signal filtered by $G_\textrm{f1}(s)=\frac{1}{(0.1s+1)^2}$ despite the presence of unmodeled external perturbation (in \refFig{fig:disturbance_FIL_1}) and white measurement noise corrupting the system output (with noise power $P_\textrm{n}=1\times 10^{-7}$ and sampling time $T_\textrm{n}=T_\textrm{s}$). The obtained FIL simulation results are presented in \refFig{fig:Results_FIL_1}.

It can be seen that the PID controller with added equivalence transfer function $C_\mathrm{EQ2}$, provides practically the same reference tracking and disturbance rejection performances as the eADRC controller. A slight difference, visible in the control signals, is a consequence of their fixed-point FPGA implementations, because '1DOF PID+$C_\mathrm{EQ2}$' structure is realized as a serial connection of \eqref{eqn:PID_z_domain} and \eqref{eqn:C_EQ2_z_domain}, while 1DOF eADRC is realized solely based on \eqref{eqn:eADRC_z_domain}. Additionally, it is evident that the eADRC-based controllers enable better disturbance rejection performances than PID controller with $T_\textrm{f}=0.005$. An improvement in the PID disturbance rejection performance is achieved by decreasing the filter coefficient ($T_\textrm{f}=0.0003$), but in that case, a significant increase of the measurement noise sensitivity becomes evident by observing the control signals in \refFig{fig:Results_FIL_1}. 

A comparison in consumption of most significant FPGA chip resources for the considered controllers, namely the look up table (LUT), digital signal processing (DSP) block, and the flip-flop (FF) block is presented in \refTable{tab:FPGA_resourse_consumption_FIL_1}. As expected, it can be seen that the 1DOF PID controller requests the least FPGA resource, and that structure '1DOF PID+$C_\mathrm{EQ2}$' occupied  more FPGA resources compared to compact 1DOF eADRC controller, due to its realization form as a serial connection of two discrete transfer functions \eqref{eqn:PID_z_domain} and \eqref{eqn:C_EQ2_z_domain}, which requires more multiplication and addition operations than 1DOF eADRC discrete transfer function \eqref{eqn:eADRC_z_domain}. However, it should be noted that in this case, a relatively low-cost FPGA chip is used, and compared with industrial FPGA chips, has significantly less resource. Hence, increasing consumption of resources, although interesting from academic perspective, is often not considered a major problem in practical industrial applications.     

\begin{table}[htb!]
    \centering
    \caption{'Scenario I' --- resource occupancy of FPGA chip XC7A35T for the considered controllers.}
    \label{tab:FPGA_resourse_consumption_FIL_1}
    {\renewcommand{\arraystretch}{1.5}
    \begin{tabular}{ccccc}
    \hline
    FPGA resource &1DOF PID  & 1DOF eADRC & 1DOF PID+$C_\mathrm{EQ2}$ & Available                                                 \\ \hline
    LUTs & 750 (3.61\%) &1005 (4.83\%) & 2331 (11.21\%) & 20800 (100\%)   \\
    FF blocks & 144 (0.35\%) &216 (0.52\%) & 288 (0.69\%) & 41600 (100\%)   \\
    DSP blocks & 43 (47.78\%) &52 (57.78\%) & 72 (80\%) & 90 (100\%)  \\  \hline
    \end{tabular}%
    }
\end{table}

\subsection{FIL simulation of 'Scenario II'}
\label{sec:FILscenario2}

The used DC motor model is expressed as a following second-order system \cite{stankovic2018optimized}:
\begin{align}
    \ddot{y}_2 &= -\frac{L_\textrm{a}F_\textrm{e}+R_\textrm{a}J_\textrm{e}}{L_\textrm{a}J_\textrm{e}} \cdot \dot{y}_2-\frac{R_\textrm{a}F_\textrm{e}+k_{\textrm{em}}k_{\textrm{me}}}{L_\textrm{a}J_\textrm{e}} \cdot y_2(t)\notag\\
    &+ \frac{k_{\textrm{em}}}{L_\textrm{a}J_\textrm{e}} \cdot \left[u_2(t)+w_2(t)\right],
    \label{eqn:DC_motor_model}
\end{align}
the where $y_2(t)$ represents the system output (DC motor angular speed), $u_2(t)$ is the control signal, $R_\textrm{a}=8.9\Omega$ is the motor armature resistance, $L_\textrm{a}=4.5$mH is the armature inductance, $J_\textrm{e}=8e^{-5}$kgm$^2$ is the motor moment of inertia, $F_\textrm{e}=12e^{-5}$kgm$^2$s/rad is the motor viscous friction coefficient, $k_{\textrm{em}}=0.105$Nm/A is the motor torque constant, $k_{\textrm{me}}=0.105$Vs/rad is the back electromotive force constant, and $w_2(t)$ represents the user-defined external perturbation (shown in \refFig{fig:disturbance_FIL_2}). 

\begin{figure*}[htb!]
   \centering
   \includegraphics[width=.9\textwidth]{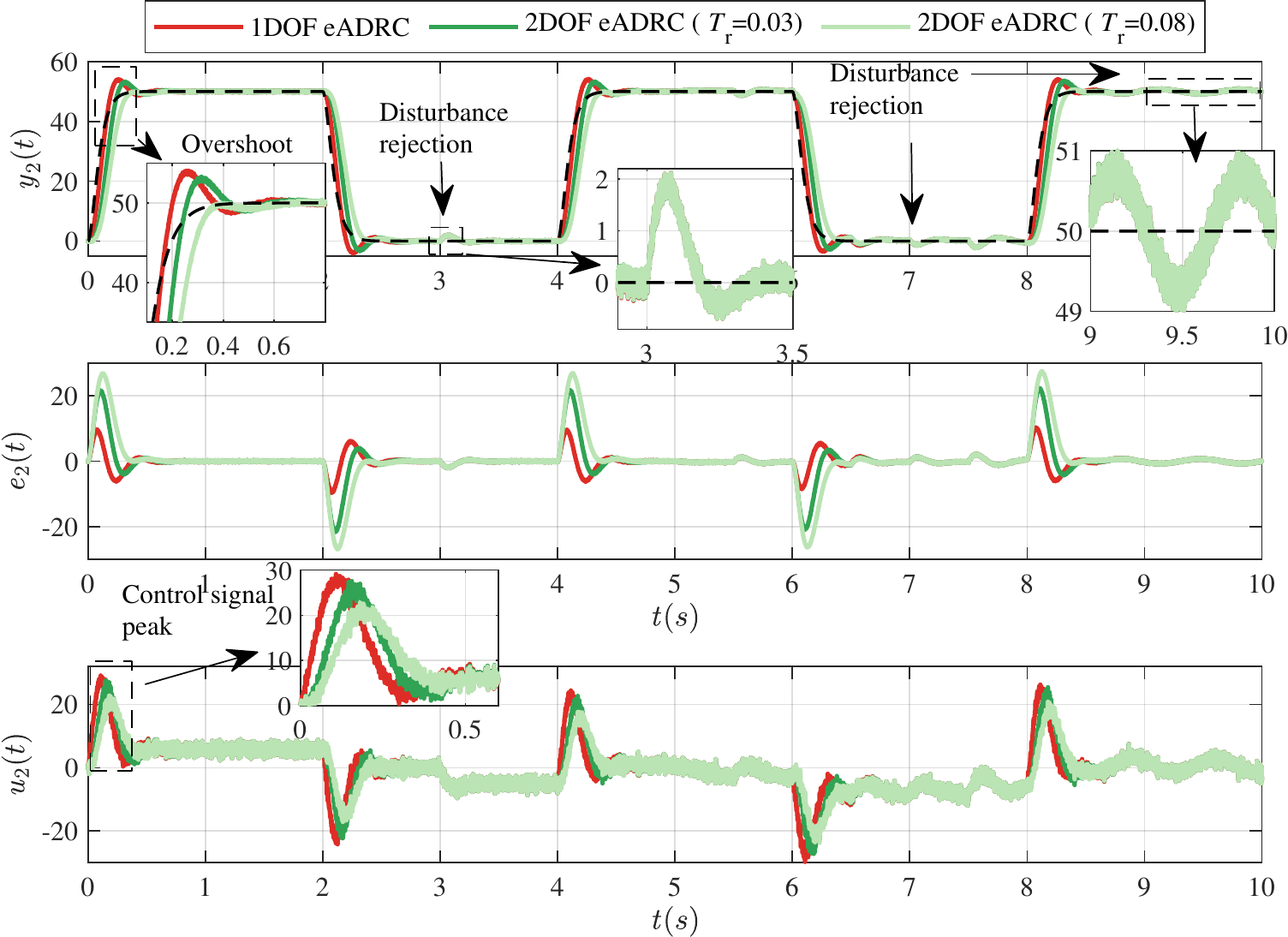}
   \caption{'Scenario II': results of FPGA-in-the-loop simulation with systems outputs (top); tracking errors (middle); control signals (bottom).}
   \label{fig:Results_FIL_2}
\end{figure*}

In the same manner as in previous scenario in \refSec{sec:FILscenario1}, the discrete forms of the considered controllers are derived by applying Euler approximation. This means that discrete 1DOF eADRC controller is realized based on \eqref{eqn:eADRC_z_domain}. In order to realize 2DOF eADRC controller, discrete form of \eqref{eqn:eADRC_pre-filter_second_order_plant}, assuming first-order filter $F_\mathrm{R}(s)=\frac{1}{T_\textrm{r}s+1}$, is obtained as:
\begin{equation}
    C_\mathrm{PF}^\mathrm{A2}(z)=\frac{H_3z^3+H_2z^2+H_1z+H_0}{Q_3z^3+Q_2z^2+Q_1z+Q_0}, 
 \label{eqn:eADRC_PF_z_domain} 
\end{equation}
with its coefficients given in Appendix~A2. Both controllers are implemented in 64-bit fixed-point format, with sampling period $T_\textrm{s}=0.001$s and tuning parameters $\omega_\textrm{CL}=50$rad/s, $k_\mathrm{ESO}=12$, and, assuming complete knowledge of the plant gain, $b_0=k_{em}/(J_\textrm{e}L_\textrm{a})$. The additional parameters related to the 2DOF structure are set to $\beta=0.6$, while two different values of the parameter $T_\textrm{r}=\left\{0.03,0.08\right\}$ are analyzed. The control objective is to track a reference trajectory being a square signal filtered by $G_\textrm{f2}(s)=\frac{1}{(0.05s+1)^2}$ despite the presence of external disturbance (in \refFig{fig:disturbance_FIL_2}) and output measurement noise (with power $P_\textrm{n}=1\times 10^{-5}$ and sampling time $T_\textrm{n}=T_\textrm{s}$). The obtained FIL simulation results are presented in \refFig{fig:Results_FIL_2}.

\Revised{From the obtained results, it is evident that the proposed 2DOF eADRC structure enables the realization of the set-point response with desired overshoot, rise time and control signal peak,  without the degradation of disturbance rejection performance and noise sensitivity. It is shown that increasing parameter $T_\textrm{r}$ in 2DOF eADRC controller decreases the system overshoot and control signal peak, but it also has the effect of increasing the rise time, i.e. increases tracking error peaks.}
Regarding FPGA resource consumption, one can see in \refTable{tab:FPGA_resourse_consumption_FIL_2} that the 2DOF eADRC controller occupied more FPGA resource than the 1DOF eADRC controller, however it does not significantly limits its implementation on low-cost FPGA hardware, such as the used XC7A35T chip.  

\begin{table}[htb!]
    \centering
    \caption{'Scenario II' --- resource occupancy of FPGA chip XC7A35T for the considered controllers.}
    \label{tab:FPGA_resourse_consumption_FIL_2}
    {\renewcommand{\arraystretch}{1.5}
    \begin{tabular}{cccc}
    \hline
    FPGA resource &1DOF eADRC  & 2DOF eADRC & Available                                                 \\ \hline
    LUTs & 1005 (4.83\%) &3358 (16.14\%) & 20800 (100\%)   \\
    FF blocks &216 (0.52\%) & 432 (1.04\%) & 41600 (100\%)   \\
    DSP blocks & 52 (57.78\%) & 69 (76.66\%) & 90 (100\%)  \\  \hline
    \end{tabular}%
    }
\end{table}

\section{Conclusions}
\label{sec:Conclusions}

In this work, we have derived formal conditions of equivalence between standard PI/PID controllers and error-based ADRC structures as well as procedures of going from one structure to another. These procedures facilitate expressing error-based ADRC schemes as standard industrial 1DOF/2DOF controllers and allow the designed controller to have desired characteristic of ADRC (i.e. strong robustness against internal and external uncertainties) while being expressed in a form familiar in engineering practice, still being dominated by PI/PID-based control schemes. These results ensure backward compatibility of future ADRC-based solutions and contribute to the adoption of active disturbance rejection-based methods in industrial practice as a viable contender to conventional PI/PID controllers.

The limitation of the proposed methodology of transitioning from error-based ADRC to PI/PID form is that it seems (at least at the moment) unique to error-based ADRC and is not directly transferable to other control schemes.

\section*{Appendix}

\subsection*{A1. Implementation \emph{crib sheet}}
\label{sec:appendixA1}

Table~\ref{tab:CheatSheetDetailed} serves as an implementation \textit{crib sheet} for controller types considered in this paper. It is prepared for the commonly used first ($n=1$) and second ($n=2$) order systems, containing ready-to-use transfer functions of PI/PID controllers and eADRC, in both 1DOF and 2DOF forms.

\begin{sidewaystable}[p]
    \centering
    \caption{Implementation \textit{crib sheet} for commonly used first ($n=1$) and second ($n=2$) order systems, containing ready-to-use transfer functions of conventional PI/PID controllers and eADRC, in 1DOF and 2DOF forms. The used acronym 'TF' stands for transfer function. Block diagram of the implementation scheme is in \refFig{fig:Visual_transition_PI/PID_eADRC}.}
    \label{tab:CheatSheetDetailed}
    {\renewcommand{\arraystretch}{2}
    \resizebox{\columnwidth}{!}{%
    \begin{tabular}{c|cc|cc}
        \toprule
        \multirow{2}{*}{Controller} & \multicolumn{2}{c|}{First-order plant model ($n=1$)} & \multicolumn{2}{c}{Second-order plant model ($n=2$)} \\ \cline{2-5} 
         & Pre-filter TF & Feedback TF  & Pre-filter TF & Feedback TF  \\ \midrule
         
        1DOF PI/PID & $C_{\textrm{PF}}^{\textrm{PI}}(s)=1$ & $C_{\textrm{FB}}^{\textrm{PI}}(s)=\frac{K_\mathrm{P}s+K_\mathrm{I}}{s}$  & $C_{\textrm{PF}}^{\textrm{PID}}(s)=1$ &$C_{\textrm{FB}}^{\textrm{PID}}(s)=\frac{K_\mathrm{D}s^2+K_\mathrm{P}s+K_\mathrm{I}}{s}F_\mathrm{Y}(s)$ \\
        1DOF eADRC & $C_{\textrm{PF}}^{\textrm{A1}}(s)=C_{\textrm{PF}}^{\textrm{PI}}(s)C_\textrm{EQ1}^{-1}=1$ & $C_{\textrm{FB}}^{\textrm{A1}}(s)=C_{\textrm{FB}}^{\textrm{PI}}(s)C_\textrm{EQ1}=\frac{1}{b_0}\cdot \frac{(k_1l_1+l_2)\cdot s+k_1l_2}{s^2+(k_1+l_2) \cdot s}$ &  $C_{\textrm{PF}}^{\textrm{A2}}(s)=C_{\textrm{PF}}^{\textrm{PID}}(s)C_\textrm{EQ2}^{-1}=1$ & $C_{\textrm{FB}}^{\textrm{A2}}(s)=C_{\textrm{FB}}^{\textrm{PID}}(s)C_\textrm{EQ2}=\frac{1}{b_0}\cdot \frac{(k_1l_1+k_2l_2+l_3)\cdot s^2+(k_1l_2+k_2l_3)\cdot s+k_1l_3}{s^3+(k_2+l_1)\cdot s^2+(l_2+k_1+l_1k_2)\cdot s}$  \\
        
        2DOF PI/PID & $C_{\textrm{PF}}^{\textrm{PI}}(s)=\frac{K_\mathrm{P}\beta s+K_\mathrm{I}}{K_\mathrm{P}s+K_\mathrm{I}}F_\mathrm{R}(s)$ & $C_{\textrm{FB}}^{\textrm{PI}}(s)=\frac{K_\mathrm{P}s+K_\mathrm{I}}{s}$ & $C_{\textrm{PF}}^{\textrm{PID}}(s)=\frac{K_\mathrm{P}\beta s+K_\mathrm{I}}{K_\mathrm{D}s^2+K_\mathrm{P}s+K_\mathrm{I}}\frac{F_\mathrm{R}(s)}{F_\mathrm{Y}(s)}$ & $C_{\textrm{FB}}^{\textrm{PID}}(s)=\frac{K_\mathrm{D}s^2+K_\mathrm{P}s+K_\mathrm{I}}{s}F_\mathrm{Y}(s)$ \\
        
        2DOF eADRC & $C_{\textrm{PF}}^{\textrm{A1}}(s)=C_{\textrm{PF}}^{\textrm{PI}}(s)C_\textrm{EQ1}^{-1}=\frac{K_\mathrm{P}\beta s+K_\mathrm{I}}{K_\mathrm{P}s+K_\mathrm{I}}\frac{F_\mathrm{R}(s)}{F_\mathrm{Y1}(s)}$ & $C_{\textrm{FB}}^{\textrm{A1}}(s)=C_{\textrm{FB}}^{\textrm{PI}}(s)C_\textrm{EQ1}=\frac{1}{b_0}\cdot \frac{(k_1l_1+l_2)\cdot s+k_1l_2}{s^2+(k_1+l_2) \cdot s}$  & $C_{\textrm{PF}}^{\textrm{A2}}(s)=C_{\textrm{PF}}^{\textrm{PID}}(s)C_\textrm{EQ2}^{-1}=\frac{K_\mathrm{P}\beta s+K_\mathrm{I}}{K_\mathrm{D}s^2+K_\mathrm{P}s+K_\mathrm{I}}\cdot \frac{F_\mathrm{R}(s)}{F_\mathrm{Y2}(s)}$ &  $C_{\textrm{FB}}^{\textrm{A2}}(s)=C_{\textrm{FB}}^{\textrm{PID}}(s)C_\textrm{EQ2}=\frac{1}{b_0}\cdot \frac{(k_1l_1+k_2l_2+l_3)\cdot s^2+(k_1l_2+k_2l_3)\cdot s+k_1l_3}{s^3+(k_2+l_1)\cdot s^2+(l_2+k_1+l_1k_2)\cdot s}$   \\[1em] 
        \bottomrule
    \end{tabular}
    }}
\end{sidewaystable}

\subsection*{A2. Coefficients of discrete transfer functions}
\label{sec:appendixA2}

\begin{itemize}

\item Coefficients of \eqref{eqn:eADRC_z_domain}:

$M_2=k_1l_1+k_2l_2+l_3$;

$M_1=-2M_2+T_\textrm{s} \cdot (k_1l_2+k_2l_3)$; 

$M_0=M_2-T_\textrm{s} \cdot \left(k_1l_2+k_2l_3\right)+T_\textrm{s}^2k_1l_3$;

$N_1=T_\textrm{s} \cdot \left(k_2+l_1\right)-2$;

$N_0=-N_1+T_\textrm{s}^2 \cdot \left(l_2+k_1+l_1k_2\right)-1$.

\item Coefficients of \eqref{eqn:C_EQ2_z_domain}:

$L_1=T_\textrm{s}T_\textrm{f}$;

$L_0=-L_1+T_\textrm{s}^2$; 

$P_2=\frac{1}{k_2l_1+l_2+k_1}$;

$P_1=P_2 \cdot \left(T_\textrm{s}k_2l_1-2\right)$;

$P_0=P_2 \cdot \left(T_\textrm{s}k_2l_1+1\right)+T_\textrm{s}^2$.

\item Coefficients of \eqref{eqn:eADRC_PF_z_domain}:

$H_3=\frac{K_\mathrm{P}\beta}{k_2l_1+l_2+k_1}$;

$H_2=-3H_3 + T_\textrm{s} \cdot \frac{K_\mathrm{P}\beta+K_\mathrm{I}(k_2+l_1)}{k_2l_1+l_2+k_1}$;

$H_1=3H_3-2T_\textrm{s} \cdot \frac{K_\mathrm{P}\beta+K_\mathrm{I}(k_2+l_1)}{k_2l_1+l_2+k_1}+T_\textrm{s}^2K_\mathrm{P}\beta \\
+T_\textrm{s}^2 \cdot \frac{K_\mathrm{I}(k_2+l+1)}{k_2l_1+l_2+k_1}$;

$H_0=-H_3+T_\textrm{s} \cdot \frac{K_\mathrm{P}\beta+K_\mathrm{I}(k_2+l_1)}{k_2l_1+l_2+k_1}-T_\textrm{s}^2K_\mathrm{P}\beta \\
-T_\textrm{s}^2 \cdot \frac{K_\mathrm{I}(k_2+l+1)}{k_2l_1+l_2+k_1}-T_\textrm{s}^3K_\mathrm{I}$;

$Q_3=K_\mathrm{D}T_\textrm{r}$;

$Q_2=-3Q_3+T_\textrm{s} \cdot \left(K_\mathrm{D}+K_\mathrm{P}T_\textrm{r}\right)$;

$Q_1=3Q_3-2T_\textrm{s} \cdot \left(K_\mathrm{D}+K_\mathrm{P}T_\textrm{r}\right)+T_\textrm{s}^2 \cdot \left(K_\mathrm{P}+K_\mathrm{I}T_\textrm{r}\right)$;

$Q_0=-Q_3+T_\textrm{s} \cdot \left(K_\mathrm{D}+K_\mathrm{P}T_\textrm{r}\right)-T_\textrm{s}^2 \cdot \left(K_\mathrm{P}+K_\mathrm{I}T_\textrm{r}\right)+T_\textrm{s}^3K_\mathrm{I}$.

\end{itemize}

\section*{Acknowledgments}

\balance

\begin{itemize}
    
    \item This work was supported by the International Foreign Expert Project Fund of Jinan University under grant number G2021199027L
    \item This work was supported by the University of Defence in Belgrade under grant VA-TT/1/21-23.
    \item There are no relevant financial or non-financial competing interests to report.
\end{itemize}

\bibliographystyle{ieeetr}
\bibliography{mybibfile}

\end{document}